\documentclass[12pt,a4paper]{article}
\pdfoutput=1
\usepackage{jheppub}
\usepackage[utf8]{inputenc}
\usepackage[T1]{fontenc}

\usepackage{tikz, pgf}  
\usetikzlibrary{patterns}
\usetikzlibrary{positioning}
\usetikzlibrary{decorations.markings}
\usetikzlibrary{intersections}
\usepackage{amsmath,physics,float}  
\usepackage{amssymb}  
\usepackage{latexsym}
\usepackage[normalem]{ulem}
\usepackage{booktabs}
\usepackage{multirow}
\usepackage{amsfonts}
\usepackage{braket}
\usepackage{mathrsfs}

\usepackage{subfig} 
\usepackage{empheq}

\usepackage[shortlabels]{enumitem}

\usepackage{pgfplots}
\pgfplotsset{compat=1.13}
\usepackage{natbib} 
\usepackage{comment}

\usepackage{bigints}
\definecolor{ashgrey}{rgb}{0.6, 0.5, 0.5}
\definecolor{ashgrey2}{rgb}{0.7, 0.73, 0.71}

\newcommand{\bg}{\begin{equation}}
\newcommand{\nd}{\end{equation}}
\newcommand{\beq}{\begin{equation}}
\newcommand{\eeq}{\end{equation}} 
\newcommand{\ef}{\rm eff}

\newcommand{\del}{\partial}
\newcommand{\lc}{\left(}
\newcommand{\rc}{\right)}

\newcommand{\ca}{\bf}

\begin{document}

\title{Glauber-Sudarshan States, Wave Functional of the Universe and the Wheeler-De Witt equation}

\author{{Suddhasattwa Brahma}$^{1}$, {Keshav Dasgupta}$^{2}$, {Fangyi Guo}$^2$, and {Bohdan Kulinich}$^{2}$}

\emailAdd{ suddhasattwa.brahma@gmail.com, keshav@hep.physics.mcgill.ca, bohdan.kulinich@mail.mcgill.ca, fguo@physics.mcgill.ca}  

\affiliation{${}^1$ Higgs Centre for Theoretical Physics, University of Edinburgh, Edinburgh, EH9 3FD, UK\\ ${}^2$ Department of Physics, McGill University, Montr\'{e}al, Qu\'{e}bec, H3A 2T8, Canada}

\begin{abstract}
{One of the pertinent question in the analysis of de Sitter as an excited state is what happens to the Glauber-Sudarshan states that are off-shell, {\it i.e.} the states that do not satisfy the Schwinger-Dyson equations. We argue that these Glauber-Sudarshan states, including the on-shell ones, are controlled by a bigger envelope wave functional namely a wave functional of the universe which surprisingly satisfies a Wheeler-De Witt equation. We provide various justification of the aforementioned identification including the determination of the emergent Hamiltonian constraint appearing in the Wheeler-De Witt equation that is satisfied by both the on- and off-shell states. Our analysis provides further evidence of why a transient four-dimensional de Sitter phase in string theory should be viewed as an excited state over a supersymmetric warped Minkowski background and not as a vacuum state.}
\end{abstract}

\maketitle


\section{Introduction: Why is quantum gravity so hard?  \label{secintro}}

Quantum gravity (QG) is fundamentally different from other non-gravitational quantum field theories (QFTs) in various respects. Some of the key differences are as follows.

\vskip.1in

\noindent $\bullet$ \textcolor{blue}{Back-reactions of the fluctuations on the background itself}: For standard QFTs, we determine the fluctuations by simply analysing the harmonic part of the Lagrangian, but the fluctuations themselves do not have to solve the full equations of motion. In QG this is not enough because the fluctuations carry energies and so they would back-react on the geometry itself thus changing it \cite{backreaction}. Therefore the correct way would be to analyze the full system, {\it i.e.} the background and the fluctuations, as an on-shell state and study the effects. However this procedure becomes exceedingly difficult if one wants to go beyond the vacuum state\footnote{Since the most probable amplitude of fluctuation over a vacuum state is the zero amplitude, the back-reaction effect vanishes. This continues to be the case even with the interacting vacuum $\vert\Omega\rangle$, but once we go away from the vacuum sector, the back-reaction effects become highly non-trivial. This is a serious problem that has not been taken carefully in the literature except in very few cases.} or even to go off-shell.

\vskip.1in

\noindent $\bullet$ \textcolor{blue}{Existence of non-local interactions in a fundamentally local theory}: In a usual QFT most of the interactions are local, and any non-local interactions are always studied in a controlled set-up (through some derivative or such expansion). In QG the story is more involved because any interaction, once expressed as a path-integral, necessitates the involvement of non-local interactions. For example, in a set-up with certain set of chosen $-$ here, for instance, the diagonal $-$ metric components, all the off-diagonal metric components are integrated away giving rise to the non-local interactions expressed using only the diagonal metric components. A priori, there is no ``heavy'' energy scale that can be associated with these integrated 
DOFs so as to have a systematic truncation of the effective action. Thus generic interactions in QG involve both local and non-local interactions of the chosen metric components\footnote{The choice of the metric components is motivated from the on-shell behavior on a given coordinate patch. Since this could potentially become ambiguous, we observe that with appropriate choice of fluxes et cetera it is always possible to keep the cross-terms between any set of sub-manifolds vanishing. We will elaborate more on this soon.}. 

\vskip.1in

\noindent $\bullet$ \textcolor{blue}{Nonexistence of a Wilsonian effective action in a generic set-up}: Usual QFT is defined at any given energy scale by an effective action that comes from integrating out the high energy modes (as well as the massive states with masses above the given scale) following a Wilsonian or an Exact Renormalization Group (ERG) procedure. For a generic set-up in QG, especially with time-dependent backgrounds, such integrating out procedure becomes impossible because of the temporal dependence of the frequencies of the fluctuations.  Additionally the UV/IR mixing more or less rules out any hope of writing an effective low energy action for the system unless we consider time-independent Minkowski background\footnote{This is important otherwise low energy supergravity over such time-independent background in string theory could not have been properly defined. The UV/IR effects manifest themselves only as providing IR cut-offs in the computations done using supergravity action at low energies.}. Put differently, such UV/IR mixing is the deep reason behind not having unitary evolution in an expanding universe with a physical UV-cutoff \cite{Cotler:2022weg}. The usual procedure of studying time-dependent background by relegating it to some kind of {\it static patch} does not help as the problem resurrects through the dynamics of the reduced density matrix corresponding to this patch.  

\vskip.1in 

\noindent $\bullet$ \textcolor{blue}{Nonexistence of free vacua and zero-interaction sectors}: Usual study of (low-energy) IR  QFT relies on a sector with decoupled oscillators which are then made to interact via some irrelevant or marginally irrelevant coupling constants. The story does get complicated when the running couplings become relevant, but then the UV picture becomes simpler. This privileged scenario is unfortunately not available in generic QG because of the non-linear nature of the gravitational interactions, although attempts have been made to study asymptotically free gravitational interactions. Thus, although we could allow weakly interacting decoupled oscillators in the far IR, at least for a time-independent background with small internal curvatures\footnote{There is a deeper level of subtlety that we are hiding here. In string theory, weak coupling ({\it i.e.} $g_s \to 0$) doesn't necessarily go hand-in-hand with weak curvatures especially when the background has some inherent temporal dependence \cite{desitter2}. Thus allowing weak string coupling and weak curvatures simultaneously require some fine tunings. We will not worry too much about it here as this will not be the main focus of the paper.}, any attempt to study QG using free vacua would likely lead to erroneous conclusions. This also means that coherent states, which are constructed from free non-interacting decoupled oscillators, do not exist in QG although they do exist in QFTs with irrelevant or marginally irrelevant coupling constants.    

\vskip.1in

\noindent $\bullet$ \textcolor{blue}{Nonexistence of positive energy minima for any potentials}: In usual QFT, supersymmerty breaking positive energy minima for any given potential are generically not forbidden. In QG, whose UV behavior is controlled by string theory\footnote{The reason is that the non-renormalizability of Einstein gravity would imply the addition of an infinite number of massive states beyond the scale ${\rm M}_s$ or ${\rm M}_p$. As far as we know, such proliferation of states can only come from the vibrational modes of closed and open strings. However an interesting approach would be to ask whether such non-renormalizability can be cured by imposing UV fixed points, or some equivalent asymptotic safety procedures. If such attempts bear positive results then the UV divergences could come under analytic control. So far unfortunately there has not been much success along this line.}, such positive energy minima for any choice of potentials are not allowed. This is sometimes presented as the statement that QG does not allow vacua with de Sitter isometries. More generally, no positive cosmological constant solutions are possible  as minima of potentials in QG although Minkowski (and possibly non scale-separated AdS) minima are allowed\footnote{For earlier studies on the instability of de Sitter, the readers may refer to \cite{polyakov}.}. This might sound contradictory to the fact that we can evade the no-go theorems \cite{GMN} by higher order curvature corrections \cite{DRS}, but the point is that such a conclusion is deeply rooted in the quantum behavior of the system that we briefly touched upon earlier (see \cite{desitter2, coherbeta, coherbeta2, borel, joydeep} for full details). This means studying four-dimensional Einstein gravity in the presence of a positive cosmological constant in the far IR will lead to inconsistent physics because such a theory does not have a good UV completion. 

\vskip.1in

\noindent $\bullet$ \textcolor{blue}{Existence of higher order interactions as trans-series expansion}: The asymptotic nature of the perturbative series necessitates the introduction of trans-series\footnote{Readers unfamiliar with trans-series may refer to \cite{transseries}.} as the correct way of expressing the interactions in any QFTs. It is then natural that we should expect trans-series expansion in QG as well. However the situation in QG is much more non-trivial. First, even before we do perturbative expansion in a path-integral set-up, higher order interactions are necessary to take care of stringy gravitational and other anomalies, including hierarchies associated with string coupling $g_s$ \cite{desitter2}. Also if we only consider on-shell degrees of freedom, then an exhaustive list of higher order perturbative expansions, for both local and non-local interactions, are at least necessary for an EFT to be defined at far IR \cite{desitter2, coherbeta, coherbeta2, borel}. Once we do perturbative expansion with the aforementioned series, then the asymptotic nature necessitates the introduction of trans-series for both local and non-local interactions \cite{maxim, joydeep}. Such trans-series expansion now involve a complicated interplay of real and complex instantons, both point-like and higher-dimensional ones, compared to a much simpler interplay of point-like instantons in any QFT set-up.

\vskip.1in

\noindent $\bullet$ \textcolor{blue}{Nonexistence of Hamiltonians and the problem of time}: In the usual study of QFT defined over a Minkowski $-$ or a time-independent $-$ background there is a unique time-like killing vector that determines the temporal evolution of the system. As such this guarantees the existence of a Hamiltonian in the system both at the classical level and as an operator in either QM or QFT. This also means that Schr\"odinger equation can be defined in QFT (and also in QM) with a wave-function that has an explicit temporal dependence\footnote{Although it is not popular to express QFT dynamics as a Schr\"odinger equation, this can nevertheless be done and results in same answer in the presence of interactions as one would have got using Feynman diagrams. In fact there are at least {\it three} kinds of Schr\"odinger equations possible in standard QFT (with irrelevant or marginally irrelevant couplings). The first one is the Schr\"odinger equation for the particles with the wave function having a probablity interpretation (see \eqref{raclrobt}). The second one is the Schr\"odinger equation for the fields with the wave functional again having a probability interpretation (see \eqref{qotom2}). Finally, the third one is a classical Schr\"odinger equation that governs the fluctuation modes over the solitonic background (see \cite{coherbeta} for a derivation). The last one clearly doesn't have any probability interpretation.}. Once we couple the system to gravity to study QG, the story changes drastically. The on-shell Hamiltonian always {\it vanishes} in such a set-up\footnote{The story is a bit more subtle here. The on-shell Hamiltonian always vanishes but this doesn't mean that the energy would always vanish. Moreover the off-shell Hamiltonian doesn't have to vanish. For an asymptotically Minkowski spacetime (in an EFT language), with a well-defined time-like killing vector, the concept of time has a definite meaning and so is the energy (despite allowing vanishing on-shell Hamiltonian).
 In fact the energy is given by the non-linear part of the Hamiltonian. However if the background is not asymptotically Minkowski, then there is no meaning of energy and both on-shell Hamiltonian and energy would vanish. We will discuss a bit more on this in section \ref{sec2.3}. \label{ohisi}}. Nonexistence of a unique time-like killing vector for generic cosmological background implies that there cannot exist a well-defined concept of energy in the system, again both at the classical {\it i.e.} on-shell, and at the quantum levels {\it i.e.} off-shell. However in all cases the Hamiltonian operator would annihilate the wave-functional of the full system, implying that the total wave functional in QG for any background $-$ be it asymptotically Minkowski or not $-$ cannot have explicit temporal dependence. (This is simply a restatement of the time-reparametrization invariance, which is the remnant of the diffeomorphism symmetry of gravity, for a homogeneous background.) Such a wave equation is called a Wheeler-De Witt equation and the solution of this equation predicts the quantum behavior of any gravitational system in QG.

\vskip.1in

\noindent  Therefore it seems on one hand, our discussion above tells us that most of the techniques used in standard QFTs unfortunately {\it cannot} be extended to QG, and any attempts to apply the well-tested rules and regulations of QFT would produce erroneous results. On the other hand, string theory provides the only possible realization of QG that can be extended to any energy scale without encountering pathologies, but the problem is that our understanding of string theory is very limited since most of the works on the subject have been restricted to the classical low energy supergravity sector defined over time-independent backgrounds\footnote{The attempt in AdS/CFT is to study {non-perturbative} QG using a dual CFT description \cite{maldacena}. In such a scenario the quantum behavior in CFT generically dualizes to the classical behavior in AdS space unless we include ${1\over {\rm N}}$ corrections in the CFT side. The latter has not been studied in much details. However since AdS space has a well-defined time-like killing vector, this is similar in some sense to the study in Minkowski spacetime. On the other hand, the dual of a cosmological evolution using a de Sitter slicing of an AdS space has its own share of problems, for example the fluxes becoming complex et cetera (see \cite{joydeep}). In de Sitter space an equivalent attempt would be to use CFT techniques to study the quantum behavior in a framework of dS/CFT \cite{strominger}. Unfortunately dS/CFT has not been convincingly demonstrated in string theory or M theory, plus {it has some features that are unique to it and not shared with AdS/CFT. For instance, unlike in AdS/CFT, the boundary CFT does not ``share'' the same notion of time with the bulk theory. Due to the cosmological horizon, it is possible that dS/CFT works in a more subtle manner analogous to the double-sided black hole in AdS space \cite{Cosmic_ER=EPR}. More on this will be presented elsewhere.}}. Thus it would seem that there is no simple way we can study time-dependent backgrounds in a UV complete theory, unless we somehow resort to globally hyperbolic\footnote{By this we mean a spacetime with well defined non-intersecting Cauchy slices that any timelike or null vector with no end points would intersect only once.} time-independent backgrounds. One immediate advantage of using such background is obvious: we can apply the full machinery of the exact renormalization group to write an EFT at low energies, plus the UV/IR correspondence is pretty harmless as it only provides an IR cut-off in the theory. Note that the alternative of using open QFT techniques \cite{vernon} to replace ERG analysis (and thus resorting back to accelerating backgrounds) cannot be applied because, one, they have not been developed for string theory or QG, and two, the backreaction effects of the fluctuations that we mentioned earlier would pose a serious hurdle to its application for the present scenario. All in all open QFT techniques are not a good alternative here. Other ways to overcome the aforementioned hurdles have been shown to be completely ineffective (see for example \cite{joydeep}). Therefore it appears that the only way out of all the aforementioned issues is to use Glauber-Sudarshan states and view de Sitter phase (or any transient accelerating spacetime) in string/M theory as an excited state over a supersymmetric warped Minkowski background \cite{coherbeta, coherbeta2, borel, joydeep}\footnote{Henceforth, unless mentioned otherwise, by a four-dimensional de Sitter space we will mean a transient four-dimensional de Sitter phase within a temporal bound, using a flat-slicing conformal coordinates, $-{1\over \sqrt{\Lambda}} < t < 0$ where $\Lambda$ is the four-dimensional cosmological constant. \label{bribich}}. 

Such a scenario immediately leads to the following question: How are the backreaction effects from the fluctuations taken into account here? To see this consider the expectation value $\langle {\bf g}_{\mu\nu}\rangle_\sigma \equiv {\bf g}^{(0)}_{\mu\nu} +\langle \delta{\bf g}_{\mu\nu}\rangle_\sigma$, where $\vert\sigma\rangle \equiv \mathbb{D}(\sigma) \vert\Omega\rangle$ is the Glauber-Sudarshan state defined using a non-unitary displacement operator $\mathbb{D}(\sigma)$ acting on the interacting vacuum $\vert\Omega\rangle$ related to the supersymmetric warped Minkowski background; and $\delta{\bf g}_{\mu\nu}$ is the metric fluctuations over this background with metric ${\bf g}^{(0)}_{\mu\nu}$ \cite{coherbeta, coherbeta2}. The analysis presented in \cite{coherbeta, coherbeta2, borel, joydeep} is done in type IIB theory, but uplifted to M-theory for technical simplifications. The backreaction of the fluctuations is easily taken care of by using the Schwinger-Dyson equation:
\bg\label{enmill}
\left\langle {\delta {\bf S}_{\rm tot}\over \delta {\bf g}^{\mu\nu}}\right\rangle_\sigma = \left\langle {\delta \over \delta {\bf g}^{\mu\nu}} \log\big(\mathbb{D}^\dagger(\sigma) \mathbb{D}(\sigma)\big)\right\rangle_\sigma, \nd
which precisely shows how the fluctuations $\delta {\bf g}_{\mu\nu}$ backreact to {\it convert} the background metric ${\bf g}^{(0)}_{\mu\nu}$ into a de Sitter metric in any required slicings. ${\bf S}_{\rm tot}$ is the total action that contains all the non-perturbative and the non-local effects, including FP ghosts and gauge fixing terms. It is expressed as a trans-series as described in detail in \cite{joydeep}. 

The equation \eqref{enmill}, and similar ones for the fluxes and the fermionic condensates, serve two purposes simultaneously: they not only show how the backreactions act back on the supersymmetric Minkowski background, but also determine the precise Glauber-Sudarshan states $\vert\sigma\rangle$ required to perform the necessary operations. However a new puzzle appears here: once the backreaction effects produce a de Sitter background, the {\it vanishing} Hamiltonian constraint should somehow automatically show up. In fact even in the Minkowski framework, although the off-shell Hamiltonian remains non-zero, the Hamiltonian operator continues to annihilate the wave functionals thus appearing to produce no dynamics for the wave functionals. This makes the question of the de Sitter Hamiltonian even more acute. Another related puzzle is the seemingly absence of the Wheeler-De Witt equation, and consequently the wave functional of the universe with four-dimensional de Sitter isometries. Where are these details hiding in our framework?

In this paper we will provide some evidence in section \ref{sec2} to suggest that the wave functional of the universe appears as an {\it envelope} wave functional over all the Glauber-Sudarshan states. The evidence include finding an infinite set of Glauber-Sudarshan states in section \ref{sec2.1}, a new renormalization procedure to construct the effective action in section \ref{sec2.2}, and finding the form of the emergent Hamiltonian in section \ref{sec2.3}. In the process we will also determine the corresponding Wheeler-De Witt equation governing the envelope wave functional in section \ref{sec2.3}, including the constraint satisfied by the emergent Hamiltonian entering the equation. We provide our conclusion in section \ref{sec3}.

\section{Evidence for the existence of a Wheeler-De Witt wave functional \label{sec2}}

First let us discuss the simple case of getting the non-relativistic Schr\"odinger equation from QFT. As this is rather well known, we will be brief. Recall that a delta function state in the configuration space of QFT is simply a classical field, albeit off-shell, in space-time (see figure 1 in \cite{borel}). On the other hand, if the delta function states in the configuration space are properly aligned, as in figure 2 in \cite{borel}, we can also get a delta function state in space. The latter is of interest to us. In other words, if $\varphi({\bf x})$ is the field operator $-$ we are taking a complex scalar field of mass $m$ for simplicity $-$ then this delta function state is simply $\vert {\bf x}\rangle = \varphi({\bf x}) \vert 0 \rangle$, where $\vert 0 \rangle$ is the local minima (with zero interactions). The corresponding Schr\"odinger wave function is then the state $\vert\psi\rangle \equiv \int d^3 {\bf x} \psi({\bf x})\vert {\bf x}\rangle$. The free Hamiltonian, corresponding to the local free vacuum $\vert 0 \rangle$ in the non-relativistic limit, is given by ${\bf H} \approx {1\over 2m}\int d^3{\bf x} \vert {\bf \nabla}\varphi\vert^2$. From here one can show by first computing $\langle{\bf x}\vert {\bf H}\vert \psi\rangle$  that:
\bg\label{raclrobt}
\left({\bf \nabla}^2 + 2m i~\partial_0\right) \psi({\bf x}, t) = 0, \nd
in the Heisenberg picture. This is the non-relativistic one-particle Schr\"odinger equation\footnote{The multi-particle generalization of the above analysis from the complex scalar field theory is straightforward and the readers can pick up details from \cite{peskin}.} with a wave function $\psi({\bf x}, t)$ that has an {\it explicit} temporal dependence. 

To extend the above computation to get a wave functional for the fields in QG is highly non-trivial. Recall that we are not trying to find a wave {\it function} for the gravitons, rather we want to find a wave {\it functional} for the gravitational fields. Clearly we need both on-shell and off-shell field configurations for this. As an example let us assume that the wave functional is given by $\Psi({\bf g}_{\mu\nu}(x), \varphi(x))$ where $x = ({\bf x}, t) \in {\bf R}^{3, 1}$ and $\varphi(x)$ is a real scalar field. For a given configuration of $\varphi(x) \equiv \varphi_1(x)$, there would be a class of on-shell metric configurations and the wave functional $\Psi({\bf g}_{\mu\nu}(x), \varphi_1(x))$ would provide the probability amplitudes for these configurations, including the probablity amplitudes of all other off-shell configurations. Changing $\varphi_1(x) \to \varphi_2(x)$ would result in another probability distribution of the on-shell and off-shell metric configurations. Combining all these together provides the wave-functional $\Psi({\bf g}_{\mu\nu}(x), \varphi(x))$. Question is, how do we quantify this in QG?

This was first successfully achieved by Hartle and Hawking \cite{HH} by using a path integral approach, or more appropriately by using a blend of path integral and canonical methods, to determine the wave functional. This was followed by a slew of works \cite{HHlater} that analyzed, and in some cases replaced, the wave functional in multiple ways. The end results of all these approaches however were similar: they provided a quantum mechanical set up to analyze the cosmological evolution of our universe. 

Unfortunately none of these computations can be extended to string theory or even to the low energy supergravity simply because, as we mentioned earlier, there is no de Sitter minima or any minima whatsoever with positive cosmological constant. The only possible way would be to resort to excited states over a local supersymmetric Minkowski minimum to find a positive cosmological constant background \cite{coherbeta, coherbeta2, borel, joydeep}. This then leads to the question of the existence of a Wheeler-De Witt wave functional: how do we know that such a wave functional exists in the configuration space of the Glauber-Sudarshan states? {A different way to pose this question would be to ask how do we know that there exists a wave functional in the space of Glauber-Sudarshan states that satisfies the gravitational constraints? In the following we provide a few hints that provide evidence for this.}

\subsection{Existence of an infinite number of off-shell Glauber-Sudarshan states \label{sec2.1}}

The Schr\"odinger equation analysis that we did above \eqref{raclrobt} can give us some hints as to how to proceed. There were three crucial steps: \textcolor{blue}{one}, finding a localized position state of the particle, \textcolor{blue}{two}, finding an envelope wave function that captures all those localized states, and \textcolor{blue}{three}, determining the dynamics of the envelope function using the field theory Hamiltonian. The natural extension of the first step is to look for all possible on and off-shell field configurations. There are two ways to go about this. First one is simple and well known. Take for example a configuration given in figure 1 of \cite{borel}. This is in general a non-trivial off-shell classical field configuration\footnote{For QFT this is definitely true. However in the presence of gravity one needs to be more careful because any off-shell configuration can become on-shell with a different choice of the matter contents ({\it i.e.} with a different choice of the energy-momentum tensor). Thus the classically accessible regions for gravity is much bigger. However for a given choice of the matter configuration there is a definite meaning to the off-shell configurations of the metric components. These are the off-shell configurations being alluded to here. \label{bokpain}}. As in the QM example earlier, such a configuration will immediately spread in the configuration space implying that an envelope wave functional governing the behavior of the system will necessarily have an explicit time dependence\footnote{This is a bit more subtle than what it appears on the face value because of the vanishing Hamiltonian constraints. We will elaborate more on this in section \ref{sec2.3}.}.
The wave equation of the system can be easily derived using Feynman path integral, {\it but this is not what we are after.} What we are after is the {\it second} way alluded to above. In the second way we again consider generic off-shell field configurations but now each of these configurations have small quantum widths which prevent them from spreading in the configuration space. All these field configurations are controlled by the Glauber-Sudarshan states $\vert\sigma\rangle$, and they take the following form \cite{coherbeta, coherbeta2, borel, joydeep}:

{\footnotesize
\bg\label{bankmey}
\langle {\bf g}_{\mu\nu}\rangle_{\sigma}  = {\bf g}^{(0)}_{\mu\nu}({\bf x}, y, w) + {1\over g^{1/\alpha}}\left[\int_0^\infty d{\rm S} ~{\rm exp}\left(-{{\rm S}\over g^{1/\alpha}}\right) {1\over 1 - {f}_{\rm max} {\rm S}^{\alpha}}\right]_{\rm P.V} \int_{k_{\rm IR}}^\mu d^{11}k ~{\sigma_{\mu\nu}(k)\over a(k)}~\psi_k(x, y, w), \nd}
which are analyzed from eleven-dimensional point of view so that 
$x^\mu \in {\bf R}^{2, 1}, {\bf x} \in {\bf R}^2$, $y^m \in {\cal M}_6$ and $ w^a \in {\mathbb{T}^2\over {\cal G}}$ where ${\cal G}$ is the group operation without fixed points. ${\bf g}^{(0)}_{\mu\nu}({\bf x}, y, w)$ is the background warped Minkowski metric. The other parameters are defined as follows: $k_{\rm IR}$ is the IR cut-off which is the remnant of the UV/IR mixing discussed earlier, $\mu$ is the energy scale beyond which all the KK and the heavy stringy modes have been integrated away, $g$ is the coupling constant defined as $g = {1\over {\rm M}_p^{+ive}}$, $a(k)$ is the graviton propagator, $\sigma_{\mu\nu}$ is the parameter associated with the Glauber-Sudarshan state $\vert\sigma\rangle$ and $\psi_k(x, y, w)$ is the fluctuation mode over the eleven-dimensional solitonic configuration ${\bf R}^{2, 1} \times {\cal M}_6 \times {\mathbb{T}^2\over {\cal G}}$. The remaining two parameters $\alpha$ and $f_{\rm max} \equiv f_{\rm max}[\sigma(k_{\rm IR}, \mu)]$ are related to the effective number of fields and the amplitude of the dominant {\it nodal digrams}. These have been derived in section 4 of \cite{joydeep} and the details of the {\it nodal diagrams} appear in \cite{borel} which the readers may look up. Finally P. V is the principal value of the integral (in square brackets) that is always positive definite irrespective of the sign of 
$f_{\rm max}$ (for a proof see section 4.5 of \cite{borel}).

One can dualize \eqref{bankmey} to type IIB and study the four-dimensional theory only where we expect a similar expression, albeit a little more involved, to appear. We will dwell very briefly on this here, but it will suffice to point out that, even from the eleven-dimensional point of view, the metric configuration will be typically off-shell. Since $\sigma_{\mu\nu}(k)$ can be any function\footnote{Albeit gauge {\it inequivalent} as we shall explain in section \ref{sec2.3}.}, there are literally an infinite number of off-shell configurations here. The on-shell metric configurations satisfy a Schwinger-Dyson equation of the form \eqref{enmill}, which may be shown to take the following simpler form (see \cite{joydeep} for a derivation):

{\scriptsize
\bg\label{eventhorizon2}
{\begin{split}
& ~~~ {\delta\hat{\bf S}_{\rm tot}(\langle{\bf \Xi}\rangle_\sigma)\over \delta \langle {\bf g}_{\mu\nu} \rangle_\sigma} ~ = ~ 0\\
& \sum_{\{\sigma_i\} \ne \sigma}\left\langle {\delta\hat{\bf S}_{\rm tot}({\bf \Xi})\over \delta{\bf g}_{\mu\nu}}\right\rangle_{(\{\sigma_i\}\vert\sigma)} + 
~\left\langle {\delta\hat{\bf S}_{\rm tot}({\bf \Xi})\over \delta  
{\bf g}_{\mu\nu}}\right\rangle_{({\bf Q}\vert\sigma)} +  \left\langle {\delta\hat{\bf S}^{({\rm g})}({\bf \Xi})\over \delta  
{\bf g}_{\mu\nu}}\right\rangle_\sigma + \left\langle {\delta\hat{\bf S}_{\rm nloc}^{({\rm g})}({\bf \Xi})\over \delta{\bf g}_{\mu\nu}}\right\rangle_\sigma = ~ \left\langle
{\delta\over \delta{\bf g}_{\mu\nu}}\log\left(\mathbb{D}_o^\dagger \mathbb{D}_o\right)\right\rangle_\sigma \\
\end{split}}
\nd}
where ${\bf \Xi} \equiv ({\bf g}_{\mu\nu}, {\bf g}_{mn}, {\bf g}_{ab}, {\bf C}_{\rm ABC}, \Psi_{\rm A}, \overline\Psi_{\rm A})$ and $({\rm A, B}) \in {\bf R}^{2, 1} \times {\cal M}_6 \times {\mathbb{T}^2\over {\cal G}}$. There is however two key differences from \eqref{enmill}: the appearance of $\hat{\bf S}_{\rm tot}$ and $\mathbb{D}_o$ instead of ${\bf S}_{\rm tot}$ and $\mathbb{D}$. The latter is the linear part of the displacement operator, and such a choice also converts parts of ${\bf S}_{\rm tot}$ to 
$\hat{\bf S}_{\rm tot}$. But the difference between ${\bf S}_{\rm tot}$ and $\hat{\bf S}_{\rm tot}$ is more non-trivial. We will discuss this a little later (see sections \ref{sec2.2} and \ref{sec2.3}). The other quantities are defined in the following way.
The parameter ${\bf Q}$ appearing as a subscript in the second equation of \eqref{eventhorizon2} comes from the resolution of identity derived in eq. (6.10) of \cite{joydeep}. The remaining two quantities are $\hat{\bf S}^{(g)}$ and $\hat{\bf S}_{\rm nloc}^{(g)}$ related to the usual ghost action and the ghost action for the non-local terms respectively (see \cite{joydeep, coherbeta2}).

The two equations in \eqref{eventhorizon2} precisely do what was promised earlier, namely capture the backreaction effects from the Glauber-Sudarshan states on the warped Minkowski background on-shell. But the other configurations from \eqref{bankmey} do not have to satisfy \eqref{eventhorizon2}. A partial plot of $\sigma_{\mu\nu}(k)$ that forms the solution of \eqref{eventhorizon2} in the gravitational sector is given in figure 16 of \cite{borel}. In the dual type IIB side using the 
Glauber-Sudarshan state $\vert\sigma'\rangle$ (which is the {\it dual} of the M-theory Glauber-Sudarshan state $\vert\sigma\rangle$ that we took above), the on-shell metric configuration would take the following ADM form:

{\footnotesize
\bg\label{omamey}
\begin{split}
ds^2 & = \langle {\bf g}_{\mu\nu}\rangle_{\sigma'} dx^\mu dx^\nu +  \langle {\bf g}_{mn}\rangle_{\sigma'} dy^m dy^n \\
& ={1\over {\rm H}^2(y)}\left[-\left({\rm N}^2(t) - {\rm N}_i(t) {\rm N}^i(t)\right)dt^2 + 2{\rm N}_i(t) dt dx^i + \gamma_{ij}({\bf x}, t) dx^i dx^j\right] + 
{\rm H}^2(y) {\bf h}_{mn}(y, t) dy^m dy^n\\
\end{split}
\nd}
where $y^m \in {\cal M}_6, x^\mu \in {\bf R}^{3, 1}, ({\bf x}, x^i) \in {\bf R}^3$ and $({\rm N}(t), {\rm N}_i(t))$ are reminiscent of the lapse and the shift functions\footnote{We denote the local temporal coordinate by $t$ in a given patch but globally there will not be a time-like killing vector.}, but because of the presence of the warp factor ${\rm H}(y)$ a simple ADM interpretation doesn't exactly go through. However from an effective four-dimensional point of view this warp factor will be invisible and then we can provide such an interpretation. The unwarped internal metric is ${\bf h}_{mn}(y, t)$ that has some time dependence. Of course the metric components in \eqref{omamey} being on-shell, the various parameters appearing therein are related to each other and to the fluxes and axio-dilaton in the full type IIB set-up. We will however consider vanishing axio-dilaton which have been shown earlier in \cite{desitter2, coherbeta, coherbeta2, borel} to solve the Schwinger-Dyson equations. Other issues like the non-existence of late-time singularities, flux quantizations, anomaly cancellations et cetera have been discussed elsewhere (see for example \cite{desitter2, coherbeta2}) so we will not elaborate further on them. The M-theory uplift of the on-shell metric configuration \eqref{omamey} takes the form:
\bg\label{vida2mey}
\begin{split}
ds^2  & = \langle {\bf g}_{\mu\nu}\rangle_{\sigma} dx^\mu dx^\nu +  \langle {\bf g}_{mn}\rangle_{\sigma'} dy^m dy^n  + \langle{\bf g}_{ab}\rangle_\sigma dw^a dw^b\\
& = {\gamma_{33}^{1/3}\over {\rm H}^{8/3}}\left[-\left({\rm N}^2 - {\rm N}_i {\rm N}^i\right)dt^2 + \left({\rm N}_i - {{\rm N}_3 \gamma_{3i}\over \gamma_{33}}\right) dt dx^i + \left(\gamma_{ij} - {\gamma_{3i}\gamma_{3j}\over \gamma_{33}}\right) dx^i dx^j\right] \\
& + \gamma_{33}^{1/3} {\rm H}^{4/3} {\bf h}_{mn} dy^m dy^n + 
{{\rm H}^{4/3}\over \gamma^{2/3}_{33}} \left(dx_3^2 + dx_{11}^2\right)\\
\end{split}
\nd
where $x^\mu \in {\bf R}^{2, 1}, x^i \in {\bf R}^2, y^m \in {\cal M}_6$ and $w^a \in {\mathbb{T}^2\over {\cal G}}$ as before. These metric configurations can be extracted from \eqref{bankmey} $-$ where we have only shown the space-time components $-$ with on-shell choices of the Glauber-Sudarshan states.
There are also non-zero G-fluxes with components ${\bf G}_{\mu\nu ab}$ whose values appear from the metric cross-terms in the type IIB side.

It is clear that there are infinite metric configurations that do not take the form \eqref{vida2mey} and would exist in the configuration space. As it happens in any QFT, existence of off-shell states always indicate the presence of a bigger envelope wave-function: it's either Schr\"odinger wave function for the point particles in the system, or wave functional for the field configurations. In the present case we are looking for the existence of a wave functional that would capture the behavior of all the on- and off-shell Glauber-Sudarshan states. Additionally, existence of the off-shell states would not only imply the presence of an emergent Hamiltonian in the system, but would also imply non-trivial renormalizations of the  coefficients of the effective action. In the following sections we will first study the emergent renormalization scheme and then discuss the emergent Hamiltonian. 

\subsection{Existence of an emergent renormalization scheme for the action parameters \label{sec2.2}}

In the previous section we pointed out the fact that the on-shell behavior is captured by an emergent action $\hat{\bf S}_{\rm tot}(\langle {\bf \Xi}\rangle_\sigma)$, shown in \eqref{eventhorizon2}, which is {\it different} from the action ${\bf S}_{\rm tot}({\bf \Xi})$ that appears in \eqref{enmill}. The differences appear from various factors and not just from the fact that the former is expressed using $\langle{\bf \Xi}\rangle_\sigma$ whereas the latter is expressed using ${\bf \Xi}$. To see this let us present the form for the two actions:
\bg\label{ortegbon}
\begin{split}
& \hat{\bf S}_{\rm tot}(\langle {\bf \Xi}\rangle_\sigma) = 
{\bf S}_{\rm kin}(\langle {\bf \Xi}\rangle_\sigma) + 
\hat{\bf S}_{\rm NP}(\langle {\bf \Xi}\rangle_\sigma) + 
\hat{\bf S}_{\rm nloc}(\langle {\bf \Xi}\rangle_\sigma)\\
& {\bf S}_{\rm tot}({\bf \Xi, \Upsilon}) =  {\bf S}_{\rm kin}({\bf \Xi}) + 
{\bf S}_{\rm NP}({\bf \Xi}) + 
{\bf S}_{\rm nloc}({\bf \Xi}) + {\bf S}^{({\rm g})}({\bf \Xi, \Upsilon}) + 
{\bf S}^{({\rm g})}_{\rm nloc}({\bf \Xi, \Upsilon})\\
\end{split}
\nd
where ${\bf \Upsilon}$ denote the ghost degrees of freedom, with the corresponding actions as
${\bf S}^{({\rm g})}({\bf \Xi}, {\bf \Upsilon})$ and 
${\bf S}^{({\rm g})}_{\rm nloc}({\bf \Xi}, {\bf \Upsilon})$ that are respectively the action and gauge fixing terms for the local non-perturbative and the non-local non-perturbative interactions respectively. Note that we do not have any explicit {\it perturbative} pieces because the action terms are expressed using trans-series so the perturbative pieces should be thought of as contributing to the zero instanton sadddles above (see \cite{joydeep} and our upcoming work \cite{hete8} for a detailed derivation of this). Looking at \eqref{ortegbon} we see that there are more than one differences:
\vskip.1in

\noindent $\bullet$ The non-perturbative piece in $\hat{\bf S}_{\rm tot}(\langle {\bf \Xi}\rangle_\sigma)$ {\it i.e.} $\hat{\bf S}_{\rm NP}(\langle {\bf \Xi}\rangle_\sigma)$,
 differs from the non-perturbative piece in ${\bf S}_{\rm tot}({\bf \Xi, \Upsilon})$
 {\it i.e.}  ${\bf S}_{\rm NP}({\bf \Xi})$.

\vskip.1in

\noindent $\bullet$ The non-local non-perturbative piece in $\hat{\bf S}_{\rm tot}(\langle {\bf \Xi}\rangle_\sigma)$ {\it i.e.} $\hat{\bf S}_{\rm nloc}(\langle {\bf \Xi}\rangle_\sigma)$,
 differs from the non-local  non-perturbative piece in ${\bf S}_{\rm tot}({\bf \Xi, \Upsilon})$
 {\it i.e.}  ${\bf S}_{\rm nloc}({\bf \Xi})$.

 \vskip.1in

 \noindent $\bullet$ The ghost and the gauge fixing terms ${\bf S}^{({\rm g})}({\bf \Xi, \Upsilon})$ and 
${\bf S}^{({\rm g})}_{\rm nloc}({\bf \Xi, \Upsilon})$ appearing above differ from the ones appearing in \eqref{eventhorizon2}.

\vskip.1in

\noindent We see that the differences lie on the forms of the action themselves, {\it i.e.} of ${\bf S}_{\rm tot}({\bf \Xi})$ and 
$\hat{\bf S}_{\rm tot}({\bf \Xi})$ even if we express them in terms of 
${\bf \Xi}$ (with similar differences in the ghost sectors). The root of these differences in fact appear from non-trivial renormalization of the coefficients of the terms of the action, thus calling for a new renormalization scheme. To see this we will take a simpler model with a real scalar field $\varphi$ in four space-time dimensions and express \eqref{enmill} in the following way: 
\bg\label{enmill2}
\left\langle {\delta {\bf S}_{\rm tot}(\varphi)\over \delta \varphi}\right\rangle_\sigma = \left\langle {\delta \over \delta \varphi} \log\big(\mathbb{D}^\dagger(\sigma) \mathbb{D}(\sigma)\big)\right\rangle_\sigma, \nd
using a Glauber-Sudarshan state $\vert\sigma\rangle = \mathbb{D}(\sigma)\vert\Omega\rangle$. We will also choose a simpler displacement operator $\mathbb{D}(\sigma) = \mathbb{D}_o(\sigma) = \int d^4x~ \sigma(x) \varphi(x)$ which is real and Hermitian but not unitary. Everything is defined over a flat Minkowski spacetime with an action given by:
\bg\label{shilorteg}
{\bf S}_{\rm tot}(\varphi) = {\bf S}_{\rm kin}(\varphi) + {\bf S}_{\rm pert}(\varphi) = \int d^4x\Big(\partial_\mu\varphi \partial^\mu \varphi + \sum_{n = 1}^\infty c_n \varphi^{2n}\Big), \nd
with $c_1 \equiv 0$ to allow for massless field. We have also taken local perturbative interactions so that the complications of the non-localities and trans-series do not appear here (we will discuss a more complete picture in the next section). Furthermore, since we are not taking derivative interactions, $c_n$ are dimensionful and so is the field $\varphi$; plus the complications of ghosts and gauge fixing do not appear here. Using the fact that:
\bg\label{scar2stan2}
\sum_\sigma {\vert\sigma\rangle \langle\sigma\vert\over \langle\sigma\vert\sigma\rangle} = \mathbb{I} + {\bf Q}(\varphi), \nd
where the form of the operator ${\bf Q}(\varphi)$ is derived in eq. (6.10) of \cite{joydeep} $-$ so that the resolution of the identity differs slightly from what we expect using coherent states $-$ the Schwinger-Dyson equation \eqref{enmill2} can be simplified to take the following form:

{\scriptsize
\bg\label{enmill3}
\left\langle {\delta {\bf S}_{\rm tot}(\varphi)\over \delta \varphi}\right\rangle_\sigma =  {\delta {\bf S}_{\rm tot}(\langle\varphi\rangle_\sigma)\over \delta \langle\varphi\rangle_\sigma} +
\sum_{\sigma' \ne \sigma}\left\langle {\delta {\bf S}_{\rm tot}(\varphi)\over \delta \varphi}\right\rangle_{(\sigma'|\sigma)} 
+ \sum_{\sigma'\ne \sigma}\left\langle {\delta {\bf S}_{\rm tot}(\varphi)\over \delta \varphi}\right\rangle_{({\bf Q}, \sigma'|\sigma)} =
\left\langle {\delta \over \delta \varphi} \log\big(\mathbb{D}^\dagger(\sigma) \mathbb{D}(\sigma)\big)\right\rangle_\sigma, \nd}
with the simpler choice of the displacement operator defined earlier. The subscripts denote intermediate off-shell states $\vert\sigma'\rangle$ with the set $\{\sigma'\} \ne \sigma$, along with the intermediate appearance of the operator ${\bf Q}(\varphi)$. (For the derivation of \eqref{enmill3} the readers may refer to section 6.3 of \cite{joydeep}.) 

What we are interested in is $\left\langle {\delta {\bf S}_{\rm tot}(\varphi)\over \delta \varphi}\right\rangle_{(\sigma'|\sigma)}$, or more appropriately $\left\langle {\delta {\bf S}_{\rm pert}(\varphi)\over \delta \varphi}\right\rangle_{(\sigma'|\sigma)}$, when $\sigma' = \sigma + \delta\sigma$, {\it i.e.} when $\sigma'$ is arbitrarily close to $\sigma$. This means we are looking at the contributions of the off-shell Glauber-Sudarshan states $\vert\sigma'\rangle$ that are arbitrarily close to the on-shell Glauber-Sudarshan state $\vert\sigma\rangle$. In other words we are looking at:

{\scriptsize
\bg\label{omachori}
\left\langle {\delta {\bf S}_{\rm pert}(\varphi)\over \delta \varphi}\right\rangle_{(\sigma'|\sigma)}  = 
\int d^4 x \sum_{n = 1}^\infty 2n c_n \langle \varphi^{2n - 1}\rangle_\sigma
 = \int d^4 x \sum_{n = 1}^\infty 2nc_n \sum_{\sigma', \sigma'',..} {\langle \sigma \vert \varphi \vert \sigma'\rangle\langle \sigma' \vert \varphi \vert \sigma''\rangle\langle \sigma'' \vert \varphi... \vert \sigma\rangle\over \langle\sigma'\vert\sigma'\rangle 
\langle\sigma''\vert\sigma''\rangle...\langle\sigma\vert\sigma\rangle} 
+ {\cal O}({\bf Q}, \{\sigma'\}), 
\nd}
where ${\cal O}({\bf Q}, \{\sigma'\})$ denote the insertions of the intermediate ${\bf Q}$ operators and the intermediate off-shell Glauber-Sudarshan states $\vert \sigma'\rangle, \vert\sigma''\rangle$ et cetera. 
To compute the effect from \eqref{omachori}, we will evaluate two representative brackets, namely $\langle\sigma\vert \varphi\vert\sigma'\rangle$ with one on-shell and one off-shell Glauber-Sudarshan states, and $\langle\sigma'\vert \varphi\vert\sigma''\rangle$
with both off-shell Glauber-Sudarshan states. The first one gives us:
\bg\label{rachelss}
\begin{split}
\langle\sigma\vert\varphi\vert\sigma'\rangle  & =  \left[1 + {\delta\sigma\over 2}{\partial f_{\rm max}\over \partial \sigma} {\partial \over \partial f_{\rm max}}\right]_{\sigma = \sigma(k_{\rm IR}, \mu)}\langle\sigma\vert\varphi \vert\sigma\rangle\\
& + {1\over 2c_n} 
\left[\int_0^\infty d{\rm S} ~{e^{-{\rm S}/c_n} \over 1 - {f}_{\rm max} {\rm S}}\right]_{\rm P.V} \int_{k_{\rm IR}}^\mu d^{11}k ~{\delta\sigma(k)\over a(k)}~ e^{-ik.x} + {\cal O}[(\delta\sigma)^2], \\
\end{split}
\nd
where we have chosen one coupling $c_n$ in \eqref{shilorteg} and not summed over $n$ to avoid introducing {\it Borel Boxes} (see section 5.2 of \cite{joydeep} for a discussion on Borel Boxes and all-order interactions). We have also ignored the momentum conservation and chosen $\alpha = 1$ in \eqref{bankmey}. The other parameter $f_{\rm max}$ is defined as $f_{\rm max} = f_{\rm max}[\sigma(k_{\rm IR}, \mu)]$. In a similar vein $\langle\sigma'\vert\varphi\vert\sigma''\rangle$ for the off-shell states $\vert\sigma'\rangle$ and $\vert\sigma''\rangle$ takes the following form:
\bg\label{rachelss2}
\begin{split}
\langle\sigma'\vert\varphi\vert\sigma''\rangle  & =  \left[1 + \sum_{i = 1}^2{\delta\sigma_i\over 2}{\partial f_{\rm max}\over \partial \sigma} {\partial \over \partial f_{\rm max}}\right]^{\sigma = \sigma(k_{\rm IR}, \mu)}_{\sigma_i = \sigma_i(k_{\rm IR}, \mu)}\langle\sigma\vert\varphi \vert\sigma\rangle\\
& + {1\over 2c_n} 
\left[\int_0^\infty d{\rm S} ~{e^{-{\rm S}/c_n} \over 1 - {f}_{\rm max} {\rm S}}\right]_{\rm P.V} \int_{k_{\rm IR}}^\mu d^{11}k \sum_{i = 1}^2{\delta\sigma_i(k)\over a(k)}~ e^{-ik.x} + {\cal O}[(\delta\sigma_i)^2], \\
\end{split}
\nd
where we have assumed that $\sigma' = \sigma + \delta\sigma_1$ and 
$\sigma'' = \sigma + \delta\sigma_2$ with 
$f_{\rm max} = f_{\rm max}[\sigma(k_{\rm IR}, \mu)]$ as before. The results from \eqref{rachelss} and \eqref{rachelss2} are essentially the same, and once we plug them in \eqref{omachori} it is easy to see that 
$\left\langle {\delta {\bf S}_{\rm pert}(\varphi)\over \delta \varphi}\right\rangle_{(\sigma + \{\delta\sigma\}|\sigma)}$ simply {\it renormalizes} the first term in \eqref{enmill3}, {\it i.e.} the term 
${\delta {\bf S}_{\rm tot}(\langle\varphi\rangle_\sigma)\over \delta \langle\varphi\rangle_\sigma}$, as long as ${\rm mag}\left\vert{\{\delta\sigma(k_{\rm IR}, \mu)\}\over \sigma(k_{\rm IR}, \mu)}\right\vert << 1$ where ${\rm mag}$ is the magnitude. In the absence of ghosts and gauge-fixing terms, this essentially converts ${\bf S}_{\rm tot}(\langle\varphi\rangle_\sigma)$ to $\hat{\bf S}_{\rm tot}(\langle\varphi\rangle_\sigma)$. In the presence of ghosts and the gauge-fixing terms it is given in \eqref{ortegbon}. In other words:

{\footnotesize
\bg\label{samwalk}
\begin{split}
& {\bf S}_{\rm tot}(\varphi) ~\xrightarrow{\rm transf} ~{\bf S}_{\rm tot}(\langle\varphi\rangle_\sigma) ~\xrightarrow{\rm renorm}~ \hat{\bf S}_{\rm tot}(\langle\varphi\rangle_\sigma)\\
& {\bf S}_{\rm tot}({\bf \Xi, \Upsilon})~ \xrightarrow{\rm transf}~ {\bf S}_{\rm tot}(\langle{\bf \Xi}\rangle_\sigma, \langle{\bf \Upsilon}\rangle_\sigma) - {\bf S}^{(g)}(\langle{\bf \Xi}\rangle_\sigma, \langle{\bf \Upsilon}\rangle_\sigma) -{\bf S}^{(g)}_{\rm nloc}(\langle{\bf \Xi}\rangle_\sigma, \langle{\bf \Upsilon}\rangle_\sigma)~\xrightarrow{\rm renorm}~ \hat{\bf S}_{\rm tot}(\langle{\bf \Xi}\rangle_\sigma)\\    \end{split}
\nd}
where the renormalized action defined over the emergent DOFs $\langle{\bf \Xi}\rangle_\sigma$ is {\it independent} of the ghosts and the gauge-fixing terms as evident from \eqref{eventhorizon2}. In fact the ghosts and the gauge-fixing terms also get renormalized in non-trivial ways which are not hard to infer (see also \cite{joydeep}). This is going to be important soon.

Before ending this section let us point out that there exists a more complicated renormalization scheme once we go away from the simple linear form of the displacement operator. For example let us consider a generic displacement operator of the form given in eq. (3.5) of \cite{joydeep}. (This can be made real by adding in the complex conjugate pieces.) The non-linear parts of $\mathbb{D}(\sigma)$ already renormalize the coefficients of the Wilsonian Effective action (or the action got by implementing the Exact Renormalization Group procedure). The off-shell pieces now change the coefficients even further so that the last step in \eqref{samwalk} would be very different from the second step. In fact adding in the non-linear parts introduce a three-step renormalization procedure. \textcolor{blue}{One}, the coefficients of the perturbative series change from the non-linear parts of $\mathbb{D}(\sigma)$. \textcolor{blue}{Two}, inserting the perturbative series in the path-integral and consequently noticing the asymptotic nature would imply Borel-\'Ecalle resummation, thus rewriting the action as a trans-series. And \textcolor{blue}{three}, adding in further renormalization of the coefficients from the neighboring off-shell Glauber-Sudarshan states following the procedure depicted earlier. This is clearly much more complicated procedure, but the end result is what we sketched in \eqref{samwalk}. How far this can be quantified will be discussed elsewhere.

\subsection{Existence of a new emergent Hamiltonian in the configuration spaces \label{sec2.3}}

Let us come back to the issue of Hamiltonian that we discussed earlier and in footnote \ref{ohisi}. In the process we will also see how to interpret the wave function when we couple the system to gravity and matter. {It was shown earlier that an explicit ADM decomposition is not of much use due to the presence of the internal directions. Thus, instead of following the standard canonical picture, we will adopt a different approach to study the wave functional.} To start, let us see how the Schr\"odinger equation appears in QM. The steps are quite well-known and can be expressed as:

{\footnotesize
\bg\label{qotom}
\begin{split}
& \psi(q_2, t) = \int_{q_1, t_1}^{q_2, t_2} {\cal D}x e^{-i{\rm S}[x]}\psi(q_1, t_1) dq_1 \\
&{\partial\psi(q_2, t)\over \partial t} = {\partial \over \partial t}
\int_{q_1, t_1}^{q_2, t_2} {\cal D}x e^{-i{\rm S}[x]}\psi(q_1, t_1) dq_1 =
i\int_{q_1, t_1}^{q_2, t_2} {\cal D}x{\rm H}(q_2, t)e^{-i{\rm S}[x]}\psi(q_1, t_1) dq_1 = i\hat{\rm H}(q_2, t) \psi(q_2, t),\\
\end{split}
\nd}
where $t = t_2 - t_1$ and in the intermediate state, the Hamiltonian ${\rm H}(q_2, t)$ is a function that is pulled outside the path-integral as an operator\footnote{In the second line of \eqref{qotom}, taking a derivative with respect to $t$ is non-trivial because it amounts to shifting the boundary of the path-integral itself. Such transformations are generally known as the Weiss transformations \cite{weiss}, and going from the second equality to the third one $-$ where we see the Hamiltonian $-$ requires a detailed analysis that we do not show here. A simpler version of this has been attempted in \cite{HH, FPI}, but for a more recent and detailed take on the computations one may refer to the interesting series of papers in \cite{feng}.}. This leads to the Schr\"odinger equation. We can also derive the same equation from QFT as shown in \eqref{raclrobt}. We should, in the process, also note the quantum mechanical identity:
\bg\label{samthisandwch}
\int {\cal D}x  {\delta {\rm S}[x]\over \delta x} e^{-i{\rm S}[x]}\psi(q_1, t_1) dq_1 = \left\langle {\delta {\rm S}[x]\over \delta x} \right\rangle = 0, \nd
which is always satisfied whether or not the system has a classical limit. The temporal evolution in QM is important and because of that the system has eigenstates with distinctive energies. In QFT the story is similar and may be represented in the following way:

{\scriptsize
\bg\label{qotom2}
\begin{split}
& \Psi_{\rm M}(\varphi_2, t) = \int_{\varphi_1, t_1}^{\varphi_2, t_2} {\cal D}\varphi e^{-i{\rm S}[\varphi]}\Psi_{\rm M}(\varphi_1, t_1) {\cal D}\varphi_1 \\
&{\partial\Psi_{\rm M}(\varphi_2, t)\over \partial t} = {\partial \over \partial t}\int_{\varphi_1, t_1}^{\varphi_2, t_2} {\cal D}\varphi e^{-i{\rm S}[\varphi]}\Psi_{\rm M}(\varphi_1, t_1) {\cal D}\varphi_1 =
i\int_{\varphi_1, t_1}^{\varphi_2, t_2} {\cal D}\varphi{\rm H}(\varphi_2, t)e^{-i{\rm S}[\varphi]}\Psi_{\rm M}(\varphi_1, t_1) {\cal D}\varphi_1 = i\hat{\rm H}(\varphi_2, t) \Psi_{\rm M}(\varphi_2, t),\\
\end{split}
\nd}
implying a similar Schr\"odinger equation for the wave functional. Note that this is different from the wave function we derived earlier in \eqref{raclrobt} as $\Psi_{\rm M}(\varphi, t)$ now denotes a wave functional for the scalar field configurations and not a wave function for the scalar {\it particles}. The distinction is important, and \eqref{qotom2} implies the existence of eigenstates $(\vert\Omega\rangle, \vert {\rm N}\rangle)$ which are generically {\it not} the harmonic oscillator states because of non-trivial interactions. $\vert\Omega\rangle$, which is taken to be non-degenerate, is the lowest energy interacting vacuum with an energy gap from the higher excited states $\vert {\rm N}\rangle$. We also expect the following identity:
\bg\label{ekover}
\int {\cal D}\varphi {\delta {\rm S}[\varphi]\over \delta \varphi} e^{-i{\rm S}[\varphi]}\Psi_{\rm M}(\varphi_1, t_1) {\cal D}\varphi_1 = \left\langle {\delta {\rm S}[\varphi]\over \delta \varphi}\right\rangle = 0, \nd
to hold whether or not the system has a classical limit or whether or not the initial wave functional is a delta function state in the configuration state. Once we replace the scalar field by other QFT DOFs, we can reproduce the structure of QFT interactions from \eqref{ekover}. The temporal evolution of the wave functional is important, and so is the distribution of the various eigenstates. The pertinent question is what happens when we couple the system to gravity? To see this we follow the same procedure as in \eqref{qotom2} to get:

{\scriptsize
\bg\label{qotom3}
\begin{split}
& \Psi(\Xi_2, t) = \sum_{\mathbb W}\int_{\Xi_1, t_1}^{\Xi_2, t_2} {\cal D}\Xi ~e^{-i{\rm S}[\Xi]}\Psi(\Xi_1, t_1) {\cal D}\Xi_1 \\
&{\partial\Psi(\Xi_2, t)\over \partial t} =\sum_{\mathbb W}{\partial \over \partial t}\int_{\Xi_1, t_1}^{\Xi_2, t_2} {\cal D}\Xi~ e^{-i{\rm S}[\Xi]}\Psi_{\rm M}(\Xi_1, t_1) {\cal D}\Xi_1 = i \sum_{\mathbb W}\int_{\Xi_1, t_1}^{\Xi_2, t_2} {\cal D}\Xi~\textcolor{red}{{\rm H}(\Xi_2, t)}~e^{-i{\rm S}[\Xi]}\Psi(\Xi_1, t_1) {\cal D}\Xi_1\\
&~~~~~~~~~~~~~~
= i\sum_{\mathbb W}\int_{\Xi_1, t_1}^{\Xi_2, t_2}  {\cal D}\Xi~\textcolor{red}{\delta {\rm S}[\Xi]\over \delta {\bf g}^{00}}\bigg\vert_{{\bf \Xi}_2}~e^{-i{\rm S}[\Xi]}\Psi(\Xi_1, t_1) {\cal D}\Xi_1 = 0,
\end{split}
\nd}
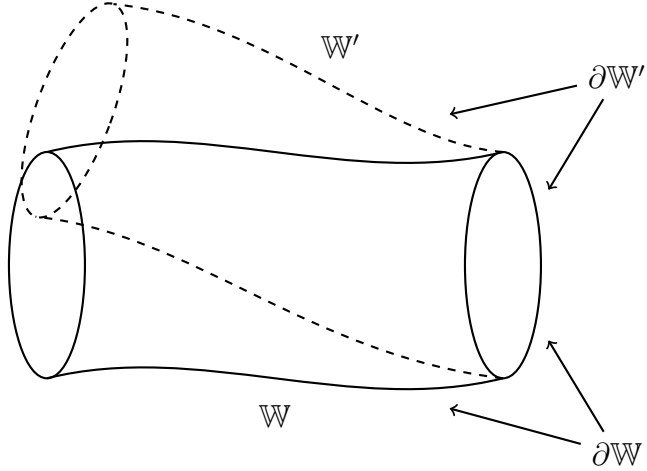
\begin{figure}
    \centering
    \begin{tikzpicture}
        \coordinate (p1) at (0, 1.5);
        \coordinate (p2) at (0, -1.5);
        \coordinate (p3) at (-6, 1.5);
        \coordinate (p4) at (-6, -1.5);
        \draw[thick, black] (0, 0) ellipse (0.5 and 1.5);
        \draw[thick, black] (-6, 0) ellipse (0.5 and 1.5);
        \draw[thick, black] (p1) .. controls (-2, 1) and (-4, 2) .. (p3);
        \draw[thick, black] (p2) .. controls (-2, -2) and (-4, -1) .. (p4);
        \node at (-3, -2) {$\mathbb{W}$};
        \begin{scope}[rotate=-20]
            \draw[thick, black, dashed] (-6, 0) ellipse (0.5 and 1.5);
            \draw[thick, black, dashed] (p1) .. controls (-2, 1) and (-4, 2) .. (-6, 1.5);
            \draw[thick, black, dashed] (p2) .. controls (-2, -2) and (-4, -1) .. (-6, -1.5);
            \node at (-3, 2) {$\mathbb{W}'$};
        \end{scope}
        \node at (1.5, -2.5) (parW) {$\partial\mathbb{W}$};
        \draw[thick, black, ->] (parW) -- (0.6, -1);
        \draw[thick, black, ->] (parW) -- (-0.7, -1.9);
        \node at (1.5, 2.5) (parW') {$\partial\mathbb{W}'$};
        \draw[thick, black, ->] (parW') -- (0.6, 1);
        \draw[thick, black, ->] (parW') -- (-0.7, 2);
    \end{tikzpicture}
    \caption{The Weiss variation \cite{weiss} that we perform here shifts the final boundary in the path-integral keeping the initial boundary coincident. The original manifold is $\mathbb{W}$ and the shifted manifold is $\mathbb{W}'$. In the end we sum over all topologies.}
    \label{rodondo}
\end{figure}
where $\Xi = ({\bf g}_{\mu\nu}, \varphi)$, from a $2+1$ dimensional point of view in M-theory or four-dimensional point of view in IIB, with the subscript denoting the initial and the final configurations; and the sum over topologies is shown explicitly (see {\bf figure \ref{rodondo}}). {The Hamiltonian is proportional to ${\delta {\rm S}[\Xi]\over \delta {\bf g}^{00}}$ since we are considering homogeneous spacetimes (which includes the case of de Sitter) for which the shift vector vanishes.}
Note the vanishing in the last step of \eqref{qotom3} because the Hamiltonian\footnote{To derive the Hamiltonian, ${\rm S}(\Xi)$ should have the appropriate Gibbons-Hawking-York boundary term. Once we allow that, the temporal variation will involve changing the boundary conditions. Including them, one can derive the results shown in the second and third equalities in the second line of \eqref{qotom3}. For more details one may refer to \cite{HH, FPI, feng}. See also \eqref{jojamez} and \eqref{omaleyla}.} is related to ${\delta {\rm S}[\Xi]\over \delta {\bf g}^{00}}$, and due to the presence of an equivalent identity like the ones from \eqref{samthisandwch} and \eqref{ekover}:
\bg\label{beatkatu}
i\sum_{\mathbb W}\int {\cal D}\Xi ~{\delta {\rm S}[\Xi]\over \delta {\bf g}^{\mu\nu}}~e^{-i{\rm S}[\Xi]}\Psi(\Xi_1, t_1) {\cal D}\Xi_1 = \left\langle  {\delta {\rm S}[\Xi]\over \delta {\bf g}^{\mu\nu}}\right\rangle = 0, \nd
the derivative of the wave functional with respect to the temporal coordinate\footnote{We are taking a globally hyperbolic spacetime with a well defined time-like killing vector to give a precise meaning to {\it time}. In the absence of a time-like killing vector one needs to resort to local clocks. Since our study of the Glauber-Sudarshan states is over asymptotically Minkowski spaces, we will not worry too much about the latter for the time being. We will come back to it soon.} vanishes. This is unfortunate because it tells us that wave functional cannot have any explicit temporal dependence, {\it i.e} $\Psi(\Xi, t) = \Psi(\Xi(t))$. How do we then reconcile with the temporal dependence of the wave functions in QFT and QM? This is a very hard question so will only be able to provide partial answers. To proceed let us carefully look at the intermediate steps in \eqref{qotom3}. First, note that the path integral is not well defined unless we add in the ghosts and the gauge fixing terms. This means for the full low energy 11 dimensional point of view, ${\rm S}(\Xi) \to {\bf S}_{\rm tot}({\bf \Xi, \Upsilon})$ defined in \eqref{ortegbon} and $\Psi(\Xi) \to {\bf \Psi}_{\rm tot}({\bf \Xi, \Upsilon})$, and the Hamiltonian will have contributions from the ghosts and the gauge fixing terms along with the contributions from ${\bf S}_{\rm NP}$ and ${\bf S}_{\rm nloc}$ (all expressed as trans-series). Secondly, comparing \eqref{enmill} and \eqref{beatkatu} we see that ${\bf S}_{\rm tot}({\bf \Xi, \Upsilon})$ satisfies the following two conditions, that coincide for vanishing $\sigma$:
\bg\label{clodugark}
\left\langle {\delta\over \delta \check{\bf \Xi}}\left({\bf S}_{\rm tot}({\bf \Xi, \Upsilon}) - {\rm log}~\mathbb{D}^\dagger({\bf \Xi}, \sigma) \mathbb{D}({\bf \Xi}, \sigma)\right)\right\rangle_\sigma = 
\left\langle {\delta{\bf S}_{\rm tot}({\bf \Xi, \Upsilon})\over \delta \check{\bf\Xi}}\right\rangle = 0, \nd
where $\check{\bf \Xi} = ({\bf \Xi, \Upsilon})$; with the former $-$ defined over the Glauber-Sudarshan states $-$ leads to the set of EOMs in \eqref{eventhorizon2}, and the latter $-$ defined over the supersymmetric warped Minkowski background $-$ leads to 
the Hamiltonian that annihilates ${\bf \Psi}({\bf \Xi, \Upsilon})$. This in turn, may be formally\footnote{{\it Formally}, because the ordering of the operators is still an unsolved problem here.} expressed as the following operation: 

{\scriptsize
\bg\label{leylaada}
\begin{split}
\hat{\bf H}_{\rm tot}(\hat{\bf \Xi}, \hat{\bf \Upsilon}) \vert{\bf \Psi}_{\rm tot}({\bf \Xi, \Upsilon})\rangle & = {\delta \hat{\bf S}_{\rm tot}(\hat{\bf \Xi}, \hat{\bf \Upsilon})\over \delta\hat{\bf g}^{00}}\vert {\bf \Psi}_{\rm tot}({\bf \Xi, \Upsilon})\rangle\\
& = \bigg(\hat{\bf H}_{\rm grav}(\hat{\bf \Xi})+ \hat{\bf H}_{\rm matter}(\hat{\bf \Xi}) + 
\hat{\bf H}_{\rm mixed}(\hat{\bf \Xi}) + \hat{\bf H}_{\rm ghost}(\hat{\bf \Xi}, \hat{\bf \Upsilon})+ \hat{\bf H}_{\rm gf} (\hat{\bf \Xi})\bigg)\vert {\bf \Psi}_{\rm tot}({\bf \Xi, \Upsilon})\rangle = 0, 
\end{split}
\nd}
where it should be clear that the Hamiltonian, action and other parameters are all operators and should not be confused with $\hat{\bf S}_{\rm tot}$ defined in \eqref{ortegbon} (to avoid such confusions we will henceforth not put hats on the operators). The Hamiltonian associated with matter ${\bf H}_{\rm matter}$ may be distinguished from ${\bf H}_{\rm mixed}$ by going to a space-time with vanishing curvature tensors where the latter would vanish and the former would reduce to the matter action that we have for the usual non-gravitational theories. In a similar vein, we expect:
\bg\label{sevgi}
\begin{split}
{\partial{\bf \Psi}_{\rm tot}({\bf \Xi, \Upsilon})\over \partial t} & = 
{\partial \over \partial t} \bigg( {\bf \Psi}_{\rm grav}({\bf \Xi}) {\bf \Psi}_{\rm matter}({\bf \Xi}) {\bf \Psi}_{\rm ghost}({\bf \Xi, \Upsilon})\bigg) = 0\\
& \implies {\partial\over \partial t}~{\rm log}~ {\bf \Psi}_{\rm grav}({\bf \Xi}) + {\partial\over \partial t}~{\rm log}~ {\bf \Psi}_{\rm matter}({\bf \Xi}) +{\partial\over \partial t}~{\rm log}~ {\bf \Psi}_{\rm ghost}({\bf \Xi, \Upsilon})= 0,\\
\end{split}
\nd
implying that ${\partial\over \partial t}{\bf \Psi}_{\rm grav}({\bf \Xi})$
and ${\partial\over \partial t}{\bf \Psi}_{\rm matter}({\bf \Xi})$ may still be non-zero and, in the absence of ghosts, would be related to each other\footnote{In fact it is not necessary to decompose the matter and the gravitational wave functionals in the aforementioned way. The point is that the wave functional of the gravity $+$ matter sector could evolve with respect to the ghost sector as long as ${\bf \Psi}_{\rm tot}({\bf \Xi, \Upsilon})$ has no explicit temporal dependence.}. However as mentioned earlier, ghosts and gauge fixing terms are absolutely necessary for the path integral to make any sense in the full low energy supergravity set-up so their absence would be alarming\footnote{One can easily show that in the path integral with metric, G-fluxes and fermionic terms, it is impossible to get rid of all ghosts. Any attempt to integrate out the ghosts is a meaningless exercise because it'll bring back the unwieldy gauge fixing conditions in the path integral that we barely managed to control by introducing the ghosts themselves! This implies that ${\bf \Psi}_{\rm tot}({\bf \Xi, \Upsilon})$ makes sense but {\it not} the wave functional involving only gravity + matter. The story is not new. We have encountered somewhat similar quantization scheme in string theory also: appearance of Faddeev-Popov ghosts in the world-sheet path-integral implies the world-sheet wave functional to have explicit ghost dependence. We can redefine the wave functional to absorb the ghosts as is done for BRST quantization. For us the story is more complicated. Ghosts appear from both gravitational and the flux sectors. For the latter the localized part of the G-fluxes in \eqref{lovrat} create non-abelian degrees of freedom as explained in \cite{dileep}. They have complicated ghost structures because of their non-abelian nature and because of their interactions with gravity. Plus the presence of higher derivative terms in \eqref{lovrat} could lead to Ostrogradsky's ghosts and other derivative ghosts. All of them therefore lead to a {\it coupled} system of ghosts, matter and gravitational DOFs as hinted in \eqref{omaleyla}. Such non-decoupling is very important for the consistency of the underlying construction such as \eqref{sevgi}. Fortunately this doesn't lead to any complications because \eqref{sevgi} is {\it not} the Wheeler-De Witt equation that we are after. The Wheeler-De Witt equation that we shall eventually get will be independent of at least the aforementioned ghosts. We would like to thank Justin Feng for helpful discussions on this and other related topics. \label{kagoney}}. The fact that they also appear in the Hamiltonian is not too hard to infer from the path integral approach. To see this we will split ${\bf S}_{\rm tot}({\bf \Xi, \Upsilon})$ as a bulk and a boundary piece in the following way:
\bg\label{jojamez}
{\bf S}_{\rm tot}({\bf \Xi, \Upsilon}) = \underbrace{\int_{\mathbb{M}} d^{n+1}x\sqrt{-{\bf g}}\left[{\cal L}_{\rm tot}({\bf \Xi, \Upsilon}) - \nabla_{\rm N} {\bf W}^{\rm N}({\bf \Xi, \Upsilon})\right]}_{{\bf S}_{\rm bulk}({\bf \Xi, \Upsilon})} + \underbrace{\int_{\del\mathbb{M}} d\Sigma^{(n)}_{\rm P} {\bf W}^{\rm P}({\bf \Xi, \Upsilon})}_{{\bf S}_{\rm bnd}({\bf \Xi, \Upsilon})},\nd
for a manifold $\mathbb{M}$ of dimension $n + 1$, where $n = 10$ for us and ${\bf W}({\bf \Xi, \Upsilon})$ is the relevant boundary piece (for the simple case when we restrict ourselves to the zero instanton sector and take only the Einstein-Hilbert term, ${\bf S}_{\rm bnd}({\bf \Xi, \Upsilon})$ will be the Gibbons-Hawking-York term \cite{GHY}). There are at least two advantages in expressing ${\bf S}_{\rm tot}({\bf \Xi, \Upsilon})$ as a bulk and a boundary piece: first, the computation that we performed in \cite{coherbeta, coherbeta2, borel, joydeep} do not receive any additional corrections from the boundary piece because it is already being included in the definition of ${\bf S}_{\rm tot}({\bf \Xi, \Upsilon})$. And secondly, the Weiss variation of the action (containing the bulk and and the boundary pieces) puts in three terms at the boundary \cite{feng}: \textcolor{blue}{one}, the total Lagrangian (including the boundary piece) for ${\bf \Xi}$ and ${\bf \Upsilon}$ with a measure given by $\delta x \cdot {\rm d}\Sigma^{(n)}$ for a $n$-dimensional boundary with a deformation $\delta x^{\rm M}$; \textcolor{blue}{two}, the contribution coming from the deformation of the boundary action ${\bf S}_{\rm bnd}({\bf \Xi, \Upsilon})$; and \textcolor{blue}{three}, the additional contribution to the boundary from the total derivative term sprouting from rewriting the bulk variation as ${\delta{\bf S}_{\rm bulk}({\bf \Xi, \Upsilon})\over \delta {\bf \Xi}}$. Together these three terms could be manipulated to take the form of the currents associated with the variations of the coordinates as well as the on-shell fields. In other words we expect \cite{feng}:

{\footnotesize
\bg\label{omaleyla}
\Delta_{\rm weiss}\Big[{\bf S}_{\rm bulk}({\bf \Xi, \Upsilon}) + 
{\bf S}_{\rm bnd}({\bf \Xi, \Upsilon})\Big] = 
\int d^{n +1} x \sqrt{-{\bf g}} ~ {\delta{\bf S}_{\rm tot}({\bf \Xi, \Upsilon})\over \delta \check{\bf \Xi}}\cdot \delta \check{\bf \Xi} + 
\int_{\del \mathbb{M}} \left(\Pi^{\rm P}_{\check{\bf \Xi}} \cdot \Delta\check{\bf \Xi} - \mathbb{H}^{\rm P}\cdot \Delta {\rm x}\right) d\Sigma^{(n)}_{\rm P}, \nd}
for the $n+1$ dimensional manifold $\mathbb{M}$ and $\mathbb{H}^{\rm P}_{\rm N}$ is the current associated with the transformation $\Delta{\rm x}^{\rm N}$ (the difference between $\delta {\rm x}^{\rm N}$ and $\Delta {\rm x}^{\rm N}$ can typically be derived following \cite{feng})\footnote{It should be clear from \eqref{jojamez} and the Weiss variation \cite{weiss} that the boundary stress-energy tensor gets the main contribution from the total action ${\bf S}_{\rm tot}({\bf \Xi, \Upsilon})$. Also since the boundary doesn't contribute when we vary with respect to $\check{\bf \Xi}$, we have replaced
${\delta{\bf S}_{\rm bulk}({\bf \Xi, \Upsilon})\over \delta \check{\bf \Xi}}$
by ${\delta{\bf S}_{\rm tot}({\bf \Xi, \Upsilon})\over \delta \check{\bf \Xi}}$ in \eqref{omaleyla}.}; and $\check{\bf \Xi} = ({\bf \Xi, \Upsilon})$. The field momenta are $\Pi^{\rm M}_{\rm N...P}$ associated with the field deformations $\Delta \check{\bf \Xi}^{\rm N...P}$; and we have identified $\mathbb{H}^0_0$ with ${\bf H}_{\rm tot}({\bf \Xi, \Upsilon})$.
The Hamiltonian can be divided in the following way:
\bg\label{selma}
\begin{split}
& {\bf H}({\bf \Xi}) = {\bf H}_{\rm grav}({\bf \Xi})+ {\bf H}_{\rm matter}({\bf \Xi}) + 
{\bf H}_{\rm mixed}({\bf \Xi})\\ 
&{\bf H}_{\rm tot}({\bf \Xi, \Upsilon}) = {\bf H}({\bf \Xi}) + {\bf H}_{\rm ghost}({\bf \Xi}, {\bf \Upsilon})+ {\bf H}_{\rm gf} ({\bf \Xi}),\\
\end{split}
\nd
where ${\bf H}_{\rm mixed}({\bf \Xi})$ contains boundary contributions;
and ${\bf \Psi}({\bf \Xi}) ={\bf \Psi}_{\rm grav}({\bf \Xi}) {\bf \Psi}_{\rm matter}({\bf \Xi})$ so that ${\bf \Psi}_{\rm tot}({\bf \Xi, \Upsilon}) = 
{\bf \Psi}({\bf \Xi}){\bf \Psi}_{\rm ghost}({\bf \Xi, \Upsilon})$. We can now express ${\bf \Psi}_{\rm tot}({\bf \Xi, \Upsilon})$ as a path integral, as in the first line in \eqref{qotom3}. Combining \eqref{omaleyla} and the second identity from \eqref{clodugark} $-$ which implies the vanishing of ${\delta{\bf S}_{\rm tot}({\bf \Xi, \Upsilon})\over \delta \check{\bf \Xi}}$ within a path integral $-$ and following the procedure outlined in \cite{feng}, we can show from the path integral expression for ${\bf \Psi}_{\rm tot}({\bf \Xi, \Upsilon})$  that:

{\scriptsize
\bg\label{sedomink}
{\rm d}{\bf \Psi}_{\rm tot}({\bf \Xi}_2, {\bf \Upsilon}_2) = \int {\cal D}{\bf \Xi} ~{\cal D}{\bf \Upsilon} \int_{\del \mathbb{M}} \left(\Pi^{\rm P}_{\check{\bf \Xi}} \cdot \Delta\check{\bf \Xi} - \mathbb{H}^{\rm P}\cdot \Delta {\rm x}\right) d\Sigma^{(n)}_{\rm P}\Big\vert_{\bf \Xi_2, \Upsilon_2}~{\rm exp}\left(-i{\bf S}_{\rm tot}[{\bf \Xi, \Upsilon}]\right) {\bf \Psi}_{\rm tot}({\bf \Xi_1, \Upsilon_1}) {\cal D}{\bf \Xi}_1 {\cal D}{\bf \Upsilon}_1, \nd}
where ${\rm d}$ is the total derivative that incorporates the Weiss variation \cite{weiss} and ${\bf S}_{\rm tot}[{\bf \Xi, \Upsilon}]$ is given in \eqref{ortegbon}. Expressing the total derivative in terms of the partial derivatives\footnote{${\rm d}{\bf \Psi}_{\rm tot}(\check{\bf \Xi}) = {\partial{\bf \Psi}_{\rm tot}(\check{\bf \Xi})\over \partial \check{\bf \Xi}} \cdot \Delta \check{\bf \Xi} +{\partial{\bf \Psi}_{\rm tot}(\check{\bf \Xi})\over \partial {\rm x}} \cdot \Delta {\rm x}$.}, immediately identifies the first line in \eqref{leylaada} with ${\partial{\bf \Psi}_{\rm tot}({\bf \Xi, \Upsilon})\over \partial t}$ where the Hamiltonian (extracted from the Hamiltonian density in \eqref{sedomink}) is given by \eqref{selma}. Now using the second identity from \eqref{clodugark}, at least in the homogeneous case, reproduces the vanishing conditions from 
\eqref{leylaada} and \eqref{sevgi} thus giving us the Wheeler-De Witt equation for the warped Minkowski set-up.

We can express the eigenstates with respect to ${\bf H}({\bf \Xi})$ to be $(\vert\Omega\rangle, \vert{\rm N}\rangle)$ where $\vert\Omega\rangle$ is the lowest energy non-degenerate interacting vacuum, and $\vert{\rm N} \rangle$ are the higher excited states with energies bigger than the energy of the vacuum state. The Glauber-Sudarshan states are constructed from this interacting vacuum as
$\vert\sigma\rangle = \mathbb{D}(\sigma) \vert\Omega\rangle$ using a non-unitary displacement operator as described earlier\footnote{It is important that we can partially isolate the sector of gravity + matter from the ghosts as in \eqref{evaada}, otherwise the displacement operator $\mathbb{D}(\sigma)$ won't be able to create an emergent ``on-shell'' configuration. A little thought will also tell us that there are no displacement operators associated with the ghosts because the ghosts only appear inside the loops and not on the external vertices.}. We can also express 
$\vert\Omega\rangle$ in terms of the local Minkowski minimum $\vert 0\rangle_{\rm loc}$ in the following way (see also \cite{joydeep}):
\bg\label{evaada}
\vert 0\rangle_{\rm loc} = c_o \vert\Omega\rangle + \sum_{{\rm N} = 1}^\infty c_{\rm N} \vert{\rm N}\rangle ~~~ \implies ~~~ 
\vert\Omega\rangle = \lim\limits_{{\rm T} \to \infty(1-i\epsilon)} {e^{-i({\bf H}({\bf \Xi}) -{\rm E}_o){\rm T}}\vert0\rangle_{\rm loc} \over
\langle \Omega \vert 0\rangle_{\rm loc}}, \nd
where ${\rm E}_o$ is the lowest eigenvalue. In deriving \eqref{evaada} we have used the fact that ${\bf H}({\bf \Xi})$ as an operator is still non-zero and only ${\bf H}_{\rm tot}({\bf \Xi, \Upsilon})$ acting on the full system annihilates the wave functional ${\bf \Psi}_{\rm tot}({\bf \Xi, \Upsilon})$.

Appearance of ${\bf H}({\bf \Xi})$ instead of ${\bf H}_{\rm tot}({\bf \Xi, \Upsilon})$ while natural for \eqref{evaada}, may appear to be somewhat problematic if we want to express for example $\langle{\bf \Xi}\rangle_\sigma$ as a path-integral as in \cite{coherbeta, coherbeta2, borel, joydeep}. However a little thinking can easily tell us that even though the path-integral would appear to allow for an action related to the Hamiltonian ${\bf H}({\bf \Xi})$, the ghosts and gauge fixing terms would automatically need to be inserted in for $\langle{\bf \Xi}\rangle_\sigma$ to make sense. This will eventually make the action to become ${\bf S}_{\rm tot}({\bf \Xi, \Upsilon})$ from \eqref{ortegbon} thus consistent with the results from \cite{coherbeta, coherbeta2, borel, joydeep} (see for example \eqref{brunekes}). Two other consistencies follow immediately from the above framework.

\vskip.1in

\noindent $\bullet$ The energy is well defined for the system because of its asymptotic Minkowski form (with a well-defined time-like direction). The reason is simple. The energy appears from the {\it linear} part of the Hamiltonian  ${\bf H}_{\rm tot}({\bf \Xi, \Upsilon})$ even though the total Hamiltonian vanishes on-shell (see for example \cite{teitelboim}). The provides the total energy from the gravitational field, the matter fields as well as the ghost fields. The off-shell Hamiltonian doesn't vanish\footnote{We do not use the word ``off-shell'' in this draft in the sense of states that are not annihilated by the gravitational constraints. Rather, this simply means off-shell trajectories in the path integral.}, yet the Hamiltonian operator annihilates the wave functional ${\bf \Psi}_{\rm tot}({\bf \Xi, \Upsilon})$. This is because of the latter's connection to the Schwinger-Dyson equation \eqref{beatkatu}.

\vskip.1in

\noindent $\bullet$ We have used the ghost fields as natural ingredients for the Minkowski spacetime relative to which, the wave functionals of the gravitational and the matter fields evolve as shown in \eqref{sevgi}.
This is motivated from the fact that the ghosts do not decouple (see footnote \ref{kagoney}) and they appear in the Hamiltonian from \eqref{omaleyla} and \eqref{selma}. (In other words $\Pi^{\rm P}_{\check{\bf \Xi}} \cdot \Delta\check{\bf \Xi}$, or any other terms, do not cancel the ghosts from the Hamiltonian in \eqref{selma}\footnote{This is guaranteed from the fact that the Hamiltonian is identified, at least in the homogeneous case, with ${\delta {\bf S}_{\rm tot}({\bf \Xi, \Upsilon})\over \delta {\bf g}^{00}}$ from \eqref{ortegbon}. Since ${\bf S}_{\rm tot}({\bf \Xi, \Upsilon})$ has ghosts, so does the Hamiltonian. This confirms \eqref{leylaada}.}.)
However this doesn't yet define a {\it clock}, for which maybe one could use the boundary value of one of the evolving matter fields as a clock since an asymptotically Minkowski spacetime allows a time-like killing vector. There could also exist mechanism akin to the  Page-Wootters prescription \cite{page} for the clock, but this would involve more complicated manipulations of the wave functional and the Hamiltonian that we leave for future to investigate. A more recent approach to the idea of a clock may be found in \cite{witten}\footnote{We thank Edward Witten for suggesting the references and for explaining the subtleties with the choice of a clock. Note that for a time-independent asymptotically Minkowski space the volume of the Cauchy slices do not change temporally, so they cannot represent a clock here.}. 

\vskip.1in

\begin{figure}
    \centering
    \begin{tikzpicture}
        \coordinate (o1) at (0, 0);
        \coordinate (p1) at (1, 0);
        \coordinate (p2) at (2.5, 0);
        \coordinate (p3) at (4, 0);
        \coordinate (p4) at (5.5, 0);
        \coordinate (p5) at (7, 0);
        \coordinate (o2) at (8, 0);
        \draw[thick] (o1) -- (o2);
        \draw[name path=l1] (p1) -- (1, 4);
        \draw[name path=l2] (p2) -- (2.5, 4);
        \draw[name path=l3] (p3) -- (4, 4);
        \draw[name path=l4] (p4) -- (5.5, 4);
        \draw[name path=l5] (p5) -- (7, 4);
        \node[below] at (p1) {${\bf \Xi}^{(1)}$};
        \node[below] at (p2) {${\bf \Xi}^{(2)}$};
        \node[below] at (p3) {${\bf \Xi}^{(3)}$};
        \node[below=0.25] at (p4) {$\cdots$};
        \draw[thick, name path=section] (0, 2) .. controls (2.5, 4) and (5.5, 1) .. (8, 1.5);
        \path[name intersections={of=section and l1,by=q1}];
        \path[name intersections={of=section and l2,by=q2}];
        \path[name intersections={of=section and l3,by=q3}];
        \path[name intersections={of=section and l4,by=q4}];
        \path[name intersections={of=section and l5,by=q5}];
        \node[circle, fill=black, inner sep=1pt] at (q1) {};
        \node[circle, fill=black, inner sep=1pt] at (q2) {};
        \node[circle, fill=black, inner sep=1pt] at (q3) {};
        \node[circle, fill=black, inner sep=1pt] at (q4) {};
        \node[circle, fill=black, inner sep=1pt] at (q5) {};
        \node at (8, 3) (g) {$\langle {\bf \Xi}\rangle_{\sigma}$};
        \draw[thick, ->] (g) -- (7.2, 1.77);
    \end{tikzpicture}
    \caption{The curved line denotes the section of the bundle entering the path integral representation of the wave functional $\Psi(\langle{\bf \Xi}\rangle_\sigma)$ in \eqref{emmwater}. The vertical lines denote the gauge orbits, or the bundles, for the various choices of the DOFs ${\bf \Xi}^{(i)}$. All these gauge redundancies enter \eqref{brunekes} which necessitates the introduction of ghosts there.}
    \label{rudeon}
\end{figure}
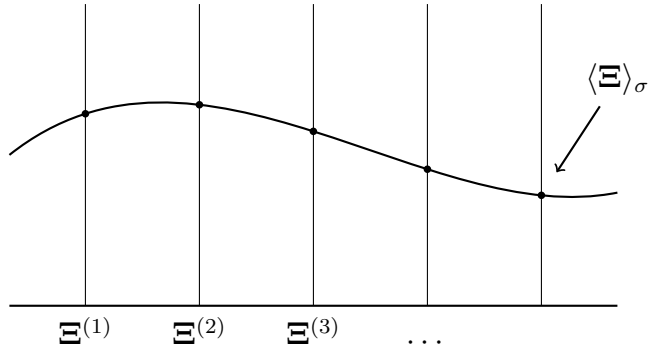

\noindent Despite the aforementioned success in explaining the dynamical behavior in the system, the idea of the presence of ghost degrees of freedom may be a bit disconcerting. Note that the finiteness of the path-integral demanded, amongst other ingredients, the presence of these ghosts, so there wasn't an option for ignoring them (see footnote \ref{kagoney}). However the beauty of the approach is that \eqref{leylaada} is {\it not} the Wheeler-De Witt equation that governs the behavior of the de Sitter background. In fact the Wheeler-De Witt equation that will eventually govern the dynamics of the Glauber-Sudarshan states is far more non-trivial. {We will call this as the {\it emergent} Wheeler-De Witt equation and the corresponding Hamiltonian entering the equation will be an emergent Hamiltonian. As is clear from the above discussion, the choice of the time-gauge for the four-dimensional de Sitter space will fix the form of the emergent Hamiltonian constraint, and thus different choices of the lapse function correspond to different Glauber-Sudarshan states. Nevertheless, these states are all related by gauge transformations (see {\bf figure \ref{rudeon}}).} The pertinent question here, of course, is how to infer the existence of such an equation in our set-up. 

The answer is not very hard to see and relies on the renormalization procedure that we outlined in the previous sub-section. The renormalization procedure, as outlined in \eqref{samwalk}, tells us how the action that governs the dynamics of a de Sitter spacetime is very different from the action that governs the dynamics in the warped Minkowski spacetime. To see this recall that the key ingredient entering any of the aforementioned action is the perturbative series:
\bg\label{lovrat}
{\bf Q}_{\rm pert}(c; {\bf \Xi}({\bf x}, y, w)) = \sum_{\{l_i\}, \{n_j\}} {{c}_{nl} \over {\rm M}_p^{\sigma_{nl}}} \left[{\bf g}^{-1}\right] \prod_{j = 0}^3 \left[\partial\right]^{n_j} \prod_{k = 1}^{60} \left({\bf R}_{\rm A_k B_k C_k D_k}\right)^{l_k} \prod_{p = 61}^{100} \left({\bf G}_{\rm A_p B_p C_p D_p}\right)^{l_p}, \nd
with only on-shell degrees of freedom (as the off-shell ones have been integrated away to allow for the non-local terms as shown in \cite{joydeep}). The perturbative piece \eqref{lovrat} is a part of a trans-series and basically contributes as the fluctuation determinant over any given instanton saddle. In other words, for every instanton saddle we have the fluctuation determinants from \eqref{lovrat} but with {\it different} set of coefficients $\{c_{nl}\}$. The instanton saddles themselves are also expressed as exponentiated integrals of \eqref{lovrat} with yet another set of $\{c_{nl}\}$ different from the ones used for the fluctuation determinants (similar story goes for the non-local terms). An explicit form for $\hat{\bf S}_{\rm tot}({\bf \Xi})$ appears in eq. (7.55) of \cite{joydeep}. Now taking in the renormalization procedure outlined in section \ref{sec2.2} we see that the key differences between ${\bf S}_{\rm tot}({\bf \Xi, \Upsilon})\big\vert_{{\bf \Upsilon} = 0}$ (with the gauge-fixing terms removed) and 
$\hat{\bf S}_{\rm tot}({\bf \Xi})$ are the set of coefficients $\{c_{nl}\}$. All the off-shell Glauber-Sudarshan states $\vert\sigma\rangle, \vert\sigma''\rangle$ et cetera located at close vicinity of the on-shell Glauber-Sudarshan state $\vert\sigma\rangle$ simply renormalize these coefficients thus converting ${\bf S}_{\rm tot}({\bf \Xi, \Upsilon})\big\vert_{{\bf \Upsilon} = 0}$ to $\hat{\bf S}_{\rm tot}({\bf \Xi})$ and then subsequently to $\hat{\bf S}_{\rm tot}(\langle{\bf \Xi}\rangle_\sigma)$! The wave functional governing all the Glauber-Sudarshan states then takes the following simple form:
\bg\label{emmwater}
\Psi(\langle{\bf \Xi}_2\rangle_\sigma) = \sum_{\cal W}\int_{\langle{\bf \Xi}_1\rangle_\sigma}^{\langle{\bf \Xi}_2\rangle_\sigma} {\cal D} {\langle{\bf \Xi}\rangle_\sigma}~{\rm exp}\left(-i
\hat{\bf S}_{\rm tot}\left[{\langle{\bf \Xi}\rangle_\sigma}\right]\right) 
\Psi({\langle{\bf \Xi}_1\rangle_\sigma}){\cal D} {\langle{\bf \Xi}_1\rangle_\sigma}, \nd
where the sum over topologies (denoted by ${\cal W}$ now) is shown explicitly. Note the {\it absence} of any ghost contributions as they have already been taken care of during the path integral representation of 
$\langle{\bf \Xi}\rangle_\sigma$. To see this recall that:
\bg\label{brunekes}
\langle{\bf \Xi}(x, y, w)\rangle_\sigma = {\int {\cal D}{\bf \Xi}~{\cal D}{\bf \Upsilon} ~{\rm exp}\Big(-i{\bf S}_{\rm tot}[{\bf \Xi, \Upsilon}]\Big) \mathbb{D}^\ast(\sigma; {\bf \Xi}) {\bf \Xi}(x, y, w) \mathbb{D}(\sigma; {\bf \Xi}) \over\int {\cal D}{\bf \Xi}~{\cal D}{\bf \Upsilon} ~{\rm exp}\Big(-i{\bf S}_{\rm tot}[{\bf \Xi, \Upsilon}]\Big) \mathbb{D}^\ast(\sigma; {\bf \Xi})\mathbb{D}(\sigma; {\bf \Xi})}, \nd
with ${\bf S}_{\rm tot}({\bf \Xi, \Upsilon})$ as given in \eqref{ortegbon}. Due to the presence of the ghosts, all the gauge orbits of ${\bf \Xi}(x, y, w)$ have already been taken care of implying that the path-integral in \eqref{emmwater} is over a section of a bundle with no ghosts appearing (see {\bf figure \ref{rudeon}}). The Wheeler-De Witt equation can now be derived by following a two-step procedure as outlined in \eqref{qotom3}.

\vskip.1in

\noindent ${\bf 1.}$ By following similar steps to generate the Schwinger-Dyson equations as discussed in \cite{joydeep}, but now applied to the wave functional \eqref{emmwater}, we get the following condition:

{\footnotesize
\bg\label{rebemoon}
\sum_{\cal W}\int_{\langle{\bf \Xi}_1\rangle_\sigma}^{\langle{\bf \Xi}_2\rangle_\sigma} {\cal D} {\langle{\bf \Xi}\rangle_\sigma}~ 
{\delta \hat{\bf S}_{\rm tot}(\langle{\bf \Xi}\rangle_\sigma) \over \delta\langle {\bf \Xi}\rangle_\sigma}{\rm exp}\left(-i
\hat{\bf S}_{\rm tot}\left[{\langle{\bf \Xi}\rangle_\sigma}\right]\right) 
\Psi({\langle{\bf \Xi}_1\rangle_\sigma}){\cal D} {\langle{\bf \Xi}_1\rangle_\sigma} \equiv\left\langle{\delta \hat{\bf S}_{\rm tot}(\langle{\bf \Xi}\rangle_\sigma) \over \delta\langle{\bf \Xi}\rangle_\sigma}\right\rangle_{\rm emergent} = 0, \nd}
from where the first emergent equation of motion from \eqref{eventhorizon2} could in principle be ascertained although the condition \eqref{rebemoon} will always be true whether or not the EOMs in \eqref{eventhorizon2} are satisfied. However the vanishing of the expectation value over the emergent spacetime in \eqref{rebemoon} is different from the condition $\left\langle{\delta \hat{\bf S}_{\rm tot}({\bf \Xi}) \over \delta{\bf \Xi}}\right\rangle_\sigma = 0$, or the two conditions in \eqref{clodugark}, or other similar conditions that were studied in \cite{coherbeta, coherbeta2, joydeep}.

\vskip.1in

\noindent ${\bf 2.}$ By performing a Weiss transformation \cite{weiss, feng} on the action 
$\hat{\bf S}_{\rm tot}(\langle{\bf \Xi}\rangle_\sigma)$, and then subsequently on the path integral \eqref{emmwater} following similar computation that led to \eqref{sedomink}, we get:
\bg\label{emmwater3}
{\partial\Psi(\langle{\bf \Xi}_2\rangle_\sigma)\over \partial t} = \sum_{\cal W}\int_{\langle{\bf \Xi}_1\rangle_\sigma}^{\langle{\bf \Xi}_2\rangle_\sigma} {\cal D} {\langle{\bf \Xi}\rangle_\sigma}~ 
{\delta \hat{\bf S}_{\rm tot}(\langle{\bf \Xi}\rangle_\sigma) \over \delta\langle {\bf g}^{00}\rangle_\sigma}\bigg\vert_{\langle{\bf \Xi}_2\rangle_\sigma}{\rm exp}\left(-i
\hat{\bf S}_{\rm tot}\left[{\langle{\bf \Xi}\rangle_\sigma}\right]\right) 
\Psi({\langle{\bf \Xi}_1\rangle_\sigma}){\cal D} {\langle{\bf \Xi}_1\rangle_\sigma}, \nd
where we have taken ${\delta \hat{\bf S}_{\rm tot}(\langle{\bf \Xi}\rangle_\sigma)\over \delta \langle{\bf g}^{00}\rangle_\sigma}$ to be proportional to the Hamiltonian since we are considering homogeneous spacetimes (which includes the case of de Sitter) for which the shift vector vanishes. Combining \eqref{rebemoon} and \eqref{emmwater3} together
immediately leads to the following equation:
\bg\label{wodewo}
{\partial\Psi(\langle{\bf \Xi}\rangle_\sigma) \over \partial t} = 
\left[{\delta \hat{\bf S}_{\rm tot}(\langle{\bf \Xi}\rangle_\sigma) \over \delta\langle {\bf g}^{00}\rangle_\sigma}\right]_{\rm oper} \Psi(\langle{\bf \Xi}\rangle_\sigma) = 0, \nd
where $[{\cal O}]_{\rm oper}$ is the operator form of ${\cal O}$, and here it represents the Hamiltonian operator.  Note that \eqref{wodewo} is exactly the kind of Wheeler-De Witt equation one would have expected for a de Sitter spacetime. This is our final answer. Let us make a few observations.

\vskip.1in

\noindent $\bullet$ Going from \eqref{emmwater} to \eqref{wodewo} has been motivated by shifting both the fields as well as the boundary conditions with the assumption that $\hat{\bf S}_{\rm tot}(\langle{\bf \Xi}\rangle_\sigma)$ allows for a boundary term (in the sense of \eqref{jojamez}). Although deriving such a boundary term is non-trivial, we know that the Weiss variation of the action would produce a {\it bulk} piece ${\delta\hat{\bf S}_{\rm bulk}(\langle{\bf \Xi}\rangle_\sigma)\over \delta\langle{\bf \Xi}\rangle_\sigma}$ as well as a {\it boundary} piece 
${\delta\hat{\bf S}_{\rm bnd}(\langle{\bf \Xi}\rangle_\sigma)\over \delta\langle{\bf \Xi}\rangle_\sigma}$
with all the currents including the Hamiltonian ${\delta \hat{\bf S}_{\rm tot}(\langle{\bf \Xi}\rangle_\sigma) \over \delta\langle {\bf g}^{00}\rangle_\sigma}$ as expected from \eqref{jojamez}. This is consistent with \cite{weiss, feng} and with the fact that we have simply split $\hat{\bf S}_{\rm tot}(\langle{\bf \Xi}\rangle_\sigma)$ into a bulk and a boundary piece. Vanishing of the two pieces within the path-integral immediately produce the vanishing on-shell Hamiltonian constraint as well as the Wheeler-De Witt equation given by \eqref{wodewo}. 

\vskip.1in

\noindent $\bullet$ The absence of ghosts in \eqref{emmwater} as well as in \eqref{wodewo} is interesting. Although reassuringly this is similar to the first reference in \cite{FPI} wherein the ghosts decoupled (see eq. (3.10) and section 4 there), it is bit more subtle now because (a) the action $\hat{\bf S}_{\rm tot}(\langle{\bf \Xi}\rangle_\sigma)$ is much more complicated than what has been taken earlier, and (b) we are not using the mini-superspace approximation. Despite this, \eqref{brunekes} justifies the absence of ghosts in $\hat{\bf S}_{\rm tot}(\langle{\bf \Xi}\rangle_\sigma)$  because all the ghosts appearing in 
${\bf S}_{\rm tot}({\bf \Xi, \Upsilon})$ essentially took care of this. In fact in the warped Minkowski case, governed by the action ${\bf S}_{\rm tot}({\bf \Xi, \Upsilon})$, the ghosts do not decouple as shown in \eqref{leylaada}, \eqref{sevgi} and \eqref{selma}\footnote{See also footnote \ref{kagoney}.}. Once we go to the emergent de Sitter case, the decoupling of ghosts doesn't always mean that there are no residual gauge symmetries left in the system. If there are remaining symmetries, the analysis will become much more non-trivial and one has to resort to finding the BRST invariant states (see the first reference in \cite{witten})\footnote{We thank Edward Witten for pointing this out.}.

\vskip.1in

\noindent $\bullet$ The problem of time in the emergent de Sitter scenario reappears. Unfortunately now, due to the decoupling of the ghosts in \eqref{wodewo}, there is no simple way we can justify the temporal evolution of the matter sector as we cannot use an equivalent formalism like \eqref{sevgi}. Interestingly however, from the warped Minkowski point of view with an {\it emergent} de Sitter, we may continue to use the temporal coordinate of Minkowski and express all dynamics from that point of view. In other words, the temporally varying volume of the Cauchy slices in the emergent de Sitter scenario can serve as a clock. This way there might be some resolution to the ``time problem'' although a more detailed study is required for us to make a definitive statement here.

\vskip.1in

\noindent $\bullet$ The interpretation of the emergent wave functional 
$\Psi(\langle{\bf \Xi}\rangle_\sigma)$ is important. It is clear that the wave functional provides the probability amplitudes for the emergent configurations $\langle{\bf \Xi}\rangle_\sigma$ and would therefore span over all the on-shell ({\it i.e.} those configurations that satisfy the Schwinger-Dyson equations \eqref{eventhorizon2}) and off-shell configurations ({\it i.e.} those configurations that do not satisfy \eqref{eventhorizon2}). For a given choice of the emergent matter configuration, there would be an unique (or a class of gauge inequivalent) emergent metric configuration(s) that would solve \eqref{eventhorizon2}. The wave functional $\Psi(\langle{\bf \Xi}\rangle_\sigma)$ provides probability amplitudes for these configurations including the ones that do not solve \eqref{eventhorizon2} for a given choice of emergent matter configuration. The configurations that do not solve \eqref{eventhorizon2} are still classically {\it allowed} regions because they can be accessed by {\it different} choices of the emergent matter configurations (see footnote \ref{bokpain}). However there exist classically {\it non-accessible} regimes where we expect the wave functional to die off exponentially whereas the wave functional should in some sense be oscillatory in the classically allowed regions. Whether the wave functional $\Psi(\langle{\bf \Xi}\rangle_\sigma)$ is peaked near certain set of configurations in the classically allowed regions is a bit challenging to find as the solution of the Wheeler-De Witt equation \eqref{wodewo} with the action $\hat{\bf S}_{\rm tot}(\langle{\bf \Xi}\rangle_\sigma)$ from \eqref{ortegbon} is highly non-trivial\footnote{In a series of interesting works \cite{suvrat}, the Wheeler-De Witt wave functional has been studied in the presence of higher order curvature corrections including the discussions on the decoupling of ghosts and the BRST invariance. It'll be worthwhile to see whether we can extend the computations of \cite{suvrat} to our set-up.}. On the other hand, if the wave functional $\Psi(\langle{\bf \Xi}\rangle_\sigma)$ starts off as sharply peaked\footnote{Assuming that it actually solves the WdW equation \eqref{wodewo}, having such a wave functional could quantify the values for $\delta\sigma$ and $\delta\sigma_i$ appearing in \eqref{rachelss} and \eqref{rachelss2} respectively that are required to implement the renormalization procedure of section \ref{sec2.2}.} over a specific Glauber-Sudarshan state in the classically allowed regions, it will continue to remain so because of the temporal invariance of the wave functional\footnote{Such temporal invariance {\it does not} prohibit the Glauber-Sudarshan states to have temporal dependence themselves. The Glauber-Sudarshan states do change with time in ways we described in \eqref{sevgi} and \eqref{evaada} (see also \cite{coherbeta, coherbeta2, borel, joydeep}), but the envelope wave functional cannot change temporally. This is what is meant for $\Psi(\langle{\bf \Xi}\rangle_\sigma)$ to have an implicit temporal dependence and not an explicit one.}. This is a more likely scenario but the question whether we can argue for the choice of the Glauber-Sudarshan state that allows for a very small four-dimensional cosmological constant, remains to be seen.

\vskip.1in

\noindent $\bullet$ The temporal domain of the validity of the Glauber-Sudarshan states is $-{1\over \sqrt{\Lambda}} < t < 0$ where $\Lambda$ is the four-dimensional cosmological constant with the emergent de Sitter spacetime being in the flat-slicing. This raises the tricky question of what happens to the wave functional for $t < -{1\over \sqrt{\Lambda}}$. From the analysis in \cite{coherbeta, coherbeta2, borel, joydeep} we know that the Glauber-Sudarshan states, as excited states over a Minkowski background, are expected to return back to the ground state beyond the aforementioned temporal domain. For the wave functional $\Psi(\langle{\bf \Xi}\rangle_\sigma)$, one possibility would be that it collapses to $\sigma = 0$ state, {\it i.e.} to the vacuum state over the supersymmetric warped Minkowski background. Such a scenario is essential for the system to also be consistent with the so-called trans-Planckian bound \cite{tcc}. How and why, or even whether, such collapse happens is not clear at this stage and more work is needed to clarify the scenario. On the other hand if the collapse really goes through, then the wave functional $\Psi(\langle{\bf \Xi}\rangle_\sigma)$ cannot quite be related to the Hartle-Hawking wave functional \cite{HH}. Unfortunately the collapsing wave functional scenario raises other conceptual issues regarding the probability conservation and the temporal invariance of the 
wave function as dictated by \eqref{wodewo}. A sharply peaked wave functional localized over non-zero $\sigma$, collapsing to another sharply peaked wave functional localized over $\sigma = 0$ at least seems to preserve probability, but not the temporal invariance. On the other hand the Hamiltonians describing $\sigma > 0$ and the $\sigma = 0$ scenarios are very different: the former appears from $\hat{\bf S}_{\rm tot}(\langle{\bf \Xi}\rangle_\sigma)$, whereas the latter appears from 
${\bf S}_{\rm tot}({\bf \Xi, \Upsilon})$. Differently speaking, once the system goes from $\sigma > 0$ to $\sigma = 0$, the Hamiltonian ${\delta \hat{\bf S}_{\rm tot}(\langle{\bf \Xi}\rangle_\sigma) \over \delta\langle {\bf g}^{00}\rangle_\sigma}$ automatically switches off and is replaced by the Hamiltonian ${\delta {\bf S}_{\rm tot}({\bf \Xi, \Upsilon}) \over \delta{\bf g}^{00}}$ allowing temporal dynamics according to \eqref{sevgi}. While this appears to provide some resolution to the issue of temporal dynamics, it is by no means a complete answer because of its failure to provide a {\it mechanism} for the collapse of the wave functional.

\vskip.1in

\noindent $\bullet$ The emergent renormalization procedure expected from having a small but non-zero width of the wave functional $\Psi(\langle{\bf \Xi}\rangle_\sigma)$ typically produces an effective action $\hat{\bf S}_{\rm tot}(\langle{\bf \Xi}\rangle_\sigma)$ that is non-Wilsonian governing the dynamics of the de Sitter excited states. In fact there are two additional ingredients contributing to the non-Wilsonian nature of the effective action: Borel-\'Ecalle resummation of the asymptotic series \cite{borel, joydeep}; and the non-linear part of the displacement operator $\mathbb{D}(\sigma)$. As mentioned earlier, such renormalization procedure distinguishes $\hat{\bf S}_{\rm tot}(\langle{\bf \Xi}\rangle_\sigma)$ $-$ even after replacing $\langle{\bf \Xi}\rangle_\sigma$ by ${\bf \Xi}$ $-$ from the equivalent piece in ${\bf S}_{\rm tot}({\bf \Xi, \Upsilon})$ with vanishing ${\bf \Upsilon}$ and vanishing gauge fixing terms. While these are all realistic theoretical predictions, probably for the first time for a non-supersymmetric background in string theory, their testability may not be feasible because of the highly sub-sub-dominant nature of the corrections. For example if we restrict $\hat{\bf S}_{\rm tot}(\langle{\bf\Xi}\rangle_\sigma)$ to say the QED sector and ask for the value of the renormalized electric charge in the energy range $k_{\rm IR} < k < \mu$, the dominant contribution to the renormalization effect would still be the radiative corrections (coming from the ERG procedure), followed by the sub-dominant instanton corrections (coming from the Borel-\'Ecalle resummation), and finally by the renormalization procedure advocated in section \ref{sec2.2} (coming from the neighboring off-shell Glauber-Sudarshan states). Despite this, it would be interesting to quantify the corrections for future testability.

\section{Discussion and conclusion \label{sec3}}

In this paper we have shown how all the Glauber-Sudarshan states, both on- and off-shell ones, are controlled by an envelope wave functional that satisfies a Wheeler-De Witt equation. This is a somewhat surprising result because the Glauber-Sudarshan states are themselves wave functional over a warped Minkowski background. The Wheeler-De Witt equation that we get in \eqref{wodewo} is an emergent equation in the sense that the de Sitter background\footnote{See footnote \ref{bribich}.} is an emergent one, described as an excited state over a warped supersymmetric Minkowski background in the far IR region of M-theory. The Hamiltonians describing the two backgrounds, the warped Minkowski and the emergent de Sitter, are also different. The difference lies in a subtle renormalization procedure that we describe in section \ref{sec2.2}. Both these Hamiltonians satisfy the necessary vanishing on-shell constraints, and they both annihilate their respective wave functionals yet, as we show in section \ref{sec2.3}, the former is capable to describe temporal variations of a subset of the matter and the gravitational wave functionals. The problem of time does appear for the emergent background, and we argue that this may be resolved by viewing the dynamics from the underlying warped Minkowski background thus providing a {\it clock} to measure temporal variations. 

The solution to the emergent Wheeler-De Witt equation \eqref{wodewo} is more challenging because the Hamiltonian appearing from the action $\hat{\bf S}_{\rm tot}(\langle{\bf \Xi}\rangle_\sigma)$ in \eqref{ortegbon} is rather complicated due to its trans-series form \cite{joydeep}. Additionally, the Hamiltonian operator would suffer from the ordering problem that we haven't resolved yet. Nevertheless certain properties of the wave functional can be ascertained from considerations related to the underlying Glauber-Sudarshan states that we dwell upon towards the end of section \ref{sec2.3}.

{The explicit solutions for four-dimensional de Sitter that have been worked out earlier in the language of Glauber-Sudarshan states has primarily been worked out for the conformal time gauge \cite{coherbeta, borel}. What our analysis highlights is that it is just as easily possible to solve for some other choice of the lapse function, equivalent to choosing, say, a cosmic time gauge for the emergent de Sitter metric in terms of Glauber-Sudarshan states. These gauge choices are made equivalent by imposing the Wheeler-De Witt constraint which, in this formalism, is an emergent Hamiltonian (constraint). Instead of applying a canonical formulation, we have shown how a Weiss variation \cite{weiss, feng} of the path integral can be applied to derive this. Our main goal was to show that solutions of the Schwinger-Dyson equations are indeed annihilated by an emergent vanishing Hamiltonian, corresponding to the resulting de Sitter space, even though all of this construction takes place over a warped Minkowski vacuum. We have mainly restricted ourselves to the homogeneous background solution for this work and in the future we plan to derive the emergent momentum constraints when dealing with small inhomegeneous fluctuations around this background. What is more, one needs to show that these emergent constraints must retain a first-class character (\`a la Dirac) and form a closed algebra \cite{Dirac-algebra}. We leave this for future work.}

All in all our analysis presented here addresses many of the pertinent questions that would typically arise in QG: the back-reaction problem; the issue of time; the Hamiltonian constraints; the Wilsonian effective action; the inherent non-localities; the interplay between the matter, gravitational and ghost sectors; the possibility of BRST ghosts; and the very existence of a four-dimensional de Sitter spacetime in string theory. 

\section*{Acknowledgements}
We would like to thank Robert Brandenberger, Joydeep Chakravarty and Alex Maloney for helpful discussions. KD thanks Justin Feng for patiently explaining the computational details of \cite{feng}; Juan Maladacena for clarifying the issue of the boundary behavior and other related topics; Savdeep Sethi and Pablo Soler for asking all the relevant questions that led to this paper; Gary Shiu, Timm Wrase and Irene Valenzuela for organising a stimulating ``The Swampland and the Landscape'' conference at the ESI, Vienna this summer; and Edward Witten for explaining the subtleties of the Wheeler-De Witt equation, the issue of time and the necessity of ghosts in the construction.

\noindent SB is supported in part by the Higgs Fellowship and by the STFC Consolidated Grant “Particle Physics at the Higgs Centre”. KD, FG and BK are supported in part by the NSERC grant.


\begin{thebibliography}{99}

\bibitem{backreaction}
V.~F.~Mukhanov, L.~R.~W.~Abramo and R.~H.~Brandenberger,
``On the Back reaction problem for gravitational perturbations,''
Phys. Rev. Lett. \textbf{78}, 1624-1627 (1997)
[arXiv:gr-qc/9609026 [gr-qc]]; 
R.~H.~Brandenberger,
``Back reaction of cosmological perturbations and the cosmological constant problem,''
[arXiv:hep-th/0210165 [hep-th]];
N.~C.~Tsamis and R.~P.~Woodard,
``Quantum gravity slows inflation,''
Nucl. Phys. B \textbf{474}, 235-248 (1996)
[arXiv:hep-ph/9602315 [hep-ph]].

\bibitem{Cotler:2022weg}
J.~Cotler and A.~Strominger,
``The Universe as a Quantum Encoder,''
[arXiv:2201.11658 [hep-th]].

\bibitem{polyakov}
A.~M.~Polyakov,
``Phase transitions and the universe,''
Sov. Phys. Usp. \textbf{25}, 187 (1982); 
``de Sitter space and eternity,''
Nucl. Phys. B \textbf{797}, 199-217 (2008)
[arXiv:0709.2899 [hep-th]]; 
``Infrared instability of the de Sitter space,''
[arXiv:1209.4135 [hep-th]].

\bibitem{GMN}
G.~W.~Gibbons,
``Aspects of supergravity theories,''
print-85-0061 (Cambridge);
J.~M.~Maldacena and C.~Nunez,
``Supergravity description of field theories on curved manifolds and a no go theorem,''
Int. J. Mod. Phys. A \textbf{16}, 822-855 (2001)
[arXiv:hep-th/0007018 [hep-th]];
G.~W.~Gibbons,
``Thoughts on tachyon cosmology,''
Class. Quant. Grav. \textbf{20}, S321-S346 (2003)
doi:10.1088/0264-9381/20/12/301
[arXiv:hep-th/0301117 [hep-th]];
K.~Dasgupta, R.~Gwyn, E.~McDonough, M.~Mia and R.~Tatar,
``de Sitter Vacua in Type IIB String Theory: Classical Solutions and Quantum Corrections,''
JHEP \textbf{07}, 054 (2014)
[arXiv:1402.5112 [hep-th]];
H.~Bernardo, S.~Brahma and M.~M.~Faruk,
``The inheritance of energy conditions: Revisiting no-go theorems in string compactifications,''
SciPost Phys. \textbf{15}, no.6, 225 (2023) [arXiv:2208.09341 [hep-th]];
J.~G.~Russo and P.~K.~Townsend,
``Time-dependent compactification to de Sitter space: a no-go theorem,''
JHEP \textbf{06}, 097 (2019)
[arXiv:1904.11967 [hep-th]].

\bibitem{DRS}
K.~Becker and M.~Becker,
``M theory on eight manifolds,''
Nucl. Phys. B \textbf{477}, 155-167 (1996)
[arXiv:hep-th/9605053 [hep-th]];
K.~Dasgupta, G.~Rajesh and S.~Sethi,
``M theory, orientifolds and G - flux,''
JHEP \textbf{08}, 023 (1999)
[arXiv:hep-th/9908088 [hep-th]];
K.~Becker and K.~Dasgupta,
``Heterotic strings with torsion,''
JHEP \textbf{11}, 006 (2002)
[arXiv:hep-th/0209077 [hep-th]]; K.~Becker, M.~Becker, K.~Dasgupta and P.~S.~Green,
``Compactifications of heterotic theory on non-K\"ahler complex manifolds. 1.,''
JHEP \textbf{04}, 007 (2003)
[arXiv:hep-th/0301161 [hep-th]]; 
G.~Lopes Cardoso, G.~Curio, G.~Dall'Agata, D.~Lust, P.~Manousselis and G.~Zoupanos,
``Non-K\"ahler string backgrounds and their five torsion classes,''
Nucl. Phys. B \textbf{652}, 5-34 (2003)
[arXiv:hep-th/0211118 [hep-th]]; 
G.~Lopes Cardoso, G.~Curio, G.~Dall'Agata and D.~Lust,
``BPS action and superpotential for heterotic string compactifications with fluxes,''
JHEP \textbf{10}, 004 (2003)
[arXiv:hep-th/0306088 [hep-th]];
K.~Becker, M.~Becker, K.~Dasgupta and S.~Prokushkin,
``Properties of heterotic vacua from superpotentials,''
Nucl. Phys. B \textbf{666}, 144-174 (2003)
[arXiv:hep-th/0304001 [hep-th]]; K.~Becker, M.~Becker, P.~S.~Green, K.~Dasgupta and E.~Sharpe,
``Compactifications of heterotic strings on non-K\"ahler complex manifolds. 2.,''
Nucl. Phys. B \textbf{678}, 19-100 (2004)
[arXiv:hep-th/0310058 [hep-th]]; M.~Becker and K.~Dasgupta,
``K\"ahler versus non-K\"ahler compactifications,''
[arXiv:hep-th/0312221 [hep-th]].


\bibitem{desitter2}
K.~Dasgupta, M.~Emelin, M.~M.~Faruk and R.~Tatar,
``de Sitter Vacua in the String Landscape,''
Nucl. Phys. B \textbf{969}, 115463 (2021)
[arXiv:1908.05288 [hep-th]];
``How a four-dimensional de Sitter solution remains outside the swampland,''
JHEP {\bf 07}, 109 (2021)
[arXiv:1911.02604 [hep-th]];
``de Sitter Vacua in the String landscape: La Petite Version,''
QTS2019
[arXiv:1911.12382 [hep-th]]; K.~Dasgupta, M.~Emelin, E.~McDonough and R.~Tatar,
``Quantum Corrections and the de Sitter Swampland Conjecture,''
JHEP \textbf{01}, 145 (2019) [arXiv:1808.07498 [hep-th]];
E.~A.~Bergshoeff, K.~Dasgupta, R.~Kallosh, A.~Van Proeyen and T.~Wrase,
``$ \overline{\mathrm{D}3} $ and dS,''
JHEP \textbf{05}, 058 (2015)
[arXiv:1502.07627 [hep-th]];
K.~Dasgupta, M.~Emelin and E.~McDonough,
``Fermions on the antibrane: Higher order interactions and spontaneously broken supersymmetry,''
Phys. Rev. D \textbf{95}, no.2, 026003 (2017)
[arXiv:1601.03409 [hep-th]].


\bibitem{coherbeta}
S.~Brahma, K.~Dasgupta and R.~Tatar,
``Four-dimensional de Sitter space is a Glauber-Sudarshan state in string theory,''
JHEP {\bf 07}, 114 (2021)
[arXiv:2007.00786 [hep-th]];\\
``de Sitter Space as a Glauber-Sudarshan State,''
JHEP {\bf 02}, 104 (2021)
[arXiv:2007.11611 [hep-th]].




\bibitem{coherbeta2}
H.~Bernardo, S.~Brahma, K.~Dasgupta, M.~M.~Faruk and R.~Tatar,
``Four-Dimensional Null Energy Condition as a Swampland Conjecture,''
Phys. Rev. Lett. \textbf{127}, no.18, 181301 (2021)
[arXiv:2107.06900 [hep-th]];
``de Sitter Space as a Glauber-Sudarshan State: II,''
Fortsch. Phys. \textbf{69}, no.11-12, 2100131 (2021)
[arXiv:2108.08365 [hep-th]];
H.~Bernardo, S.~Brahma, K.~Dasgupta and R.~Tatar,
``Crisis on Infinite Earths: Short-lived de Sitter Vacua in the String Theory Landscape,''
JHEP \textbf{04}, 037 (2021)
[arXiv:2009.04504 [hep-th]];
``Purely nonperturbative AdS vacua and the swampland,''
Phys. Rev. D \textbf{104}, no.8, 086016 (2021)
[arXiv:2104.10186 [hep-th]];
M.~M.~Faruk,
``Deriving the Gibbons-Maldacena-Nunez no-go theorem from the Raychaudhuri equation,''
Phys. Rev. D \textbf{109}, no.6, L061902 (2024)
[arXiv:2402.08805 [hep-th]].

\bibitem{borel}
S.~Brahma, K.~Dasgupta, M.~M.~Faruk, B.~Kulinich, V.~Meruliya, B.~Pym and R.~Tatar,
``Resurgence of a de Sitter Glauber-Sudarshan State: Nodal Diagrams and Borel Resummation,''
Fortsch. Phys. \textbf{71}, no.12, 2300136 (2023)
[arXiv:2211.09181 [hep-th]].

\bibitem{joydeep}
J.~Chakravarty and K.~Dasgupta,
``What if string theory has a de Sitter excited state?,''
[arXiv:2404.11680 [hep-th]].

\bibitem{transseries}
See M.~Mari\~no,
``Lectures on non-perturbative effects in large $N$ gauge theories, matrix models and strings,''
Fortsch. Phys. \textbf{62}, 455-540 (2014)
[arXiv:1206.6272 [hep-th]], and references therein;
G.~V.~Dunne and M.~\"Unsal,
``Resurgence and Trans-series in Quantum Field Theory: The CP${}^{{\rm N}-1}$ Model,''
JHEP \textbf{11}, 170 (2012), [arXiv:1210.2423 [hep-th]]; 
``Generating nonperturbative physics from perturbation theory,''
Phys. Rev. D \textbf{89}, no.4, 041701 (2014), 
[arXiv:1306.4405 [hep-th]]; 
G.~Basar, G.~V.~Dunne and M.~\"Unsal,
``Resurgence theory, ghost-instantons, and analytic continuation of path integrals,''
JHEP \textbf{10}, 041 (2013), [arXiv:1308.1108 [hep-th]];
S.~Gukov, M.~Mari\~no and P.~Putrov,
``Resurgence in complex Chern-Simons theory,''
[arXiv:1605.07615 [hep-th]];
L.~Di Pietro, M.~Mari\~no, G.~Sberveglieri and M.~Serone,
``Resurgence and 1/N Expansion in Integrable Field Theories,''
JHEP \textbf{10}, 166 (2021)
[arXiv:2108.02647 [hep-th]];
D.~Dorigoni,
``An Introduction to Resurgence, Trans-Series and Alien Calculus,''
Annals Phys. \textbf{409}, 167914 (2019)
[arXiv:1411.3585 [hep-th]].


\bibitem{maxim}
M.~Emelin,
``Effective Theories as Truncated Trans-Series and Scale Separated Compactifications,''
JHEP \textbf{11}, 144 (2020)
[arXiv:2005.11421 [hep-th]]; 
``Obstacles for dS in Supersymmetric Theories,''
PoS \textbf{CORFU2021}, 187 (2022)
[arXiv:2206.01603 [hep-th]].



\bibitem{maldacena}
J.~M.~Maldacena,
``The Large N limit of superconformal field theories and supergravity,''
Adv. Theor. Math. Phys. \textbf{2}, 231-252 (1998)
[arXiv:hep-th/9711200 [hep-th]].

\bibitem{strominger}
A.~Strominger,
``The dS / CFT correspondence,''
JHEP \textbf{10}, 034 (2001)
[arXiv:hep-th/0106113 [hep-th]].

\bibitem{Cosmic_ER=EPR}
J.~Cotler and A.~Strominger,
``Cosmic ER=EPR in dS/CFT,''
[arXiv:2302.00632 [hep-th]].

\bibitem{vernon}
R.~P.~Feynman and F.~L.~Vernon, Jr.,
``The Theory of a general quantum system interacting with a linear dissipative system,''
Annals Phys. \textbf{24}, 118-173 (1963);
C.~Agon, V.~Balasubramanian, S.~Kasko and A.~Lawrence,
``Coarse Grained Quantum Dynamics,''
Phys. Rev. D \textbf{98}, no.2, 025019 (2018)
[arXiv:1412.3148 [hep-th]].
S.~Brahma, A.~Berera and J.~Calder\'on-Figueroa,
``Quantum corrections to the primordial tensor spectrum: open EFTs \& Markovian decoupling of UV modes,''
JHEP \textbf{08}, 225 (2022)
[arXiv:2206.05797 [hep-th]].
T.~Colas, J.~Grain and V.~Vennin,
``Benchmarking the cosmological master equations,''
[arXiv:2209.01929 [hep-th]];
S.~Brahma, J.~Calder\'on-Figueroa and X.~Luo,
``Time-convolutionless cosmological master equations: Late-time resummations and decoherence for non-local kernels,''
[arXiv:2407.12091 [hep-th]].


\bibitem{peskin}
M.~E.~Peskin and D.~V.~Schroeder,
``An Introduction to quantum field theory,''
Addison-Wesley (1995);
D.~Tong,
``Lectures on the Quantum Field Theory,'' University of Cambridge Part III Mathematical Tripos (2006); 
https://www.damtp.cam.ac.uk/user/tong/qft.html

\bibitem{HH}
J.~B.~Hartle and S.~W.~Hawking,
``Wave Function of the Universe,''
Phys. Rev. D \textbf{28}, 2960-2975 (1983)

\bibitem{HHlater}
S.~W.~Hawking,
``The Quantum State of the Universe,''
Nucl. Phys. B \textbf{239}, 257 (1984);
A.~Vilenkin,
``Quantum Creation of Universes,''
Phys. Rev. D \textbf{30}, 509-511 (1984);
``Boundary Conditions in Quantum Cosmology,''
Phys. Rev. D \textbf{33}, 3560 (1986);
``Quantum Cosmology and the Initial State of the Universe,''
Phys. Rev. D \textbf{37}, 888 (1988);
``The Interpretation of the Wave Function of the Universe,''
Phys. Rev. D \textbf{39}, 1116 (1989);
A.~D.~Linde,
``Quantum creation of an inflationary universe,''
Sov. Phys. JETP \textbf{60}, 211-213 (1984); ``Quantum Creation of the Inflationary Universe,''
Lett. Nuovo Cim. \textbf{39}, 401-405 (1984).

\bibitem{hete8}
S.~Brahma, K.~Dasgupta, A.~Maji, B.~Kulinich, P.~Ramadevi and R.~Tatar,
``de Sitter and Quasi de Sitter States in ${\rm SO}(32)$ and ${\rm E}_8 \times {\rm E}_8$ Heterotic String Theories,'' {\it To Appear};
S.~Alexander, K.~Dasgupta, A.~Maji, P.~Ramadevi and R.~Tatar,
``de Sitter State in Heterotic String Theory,''
[arXiv:2303.12843 [hep-th]].

\bibitem{teitelboim}
T.~Regge and C.~Teitelboim,
``Role of Surface Integrals in the Hamiltonian Formulation of General Relativity,''
Annals Phys. \textbf{88}, 286 (1974);
S.~Deser, J.~H.~Kay and K.~S.~Stelle,
``Hamiltonian Formulation of Supergravity,''
Phys. Rev. D \textbf{16}, 2448 (1977);
S.~Deser and C.~Teitelboim,
``Supergravity Has Positive Energy,''
Phys. Rev. Lett. \textbf{39}, 249 (1977)

\bibitem{weiss}
P.~Weiss,
``On the quantization of a theory arising from a variational principle for multiple integrals with application to Born's electrodynamics,''
Proc. R. Soc. A \textbf{156}, 192 (1936); A textbook treatment of Weiss variation appears in:
N.~Mukunda and E.~C.~G.~Sudarshan,
``Classical dynamics: a modern perspective,'' Chapter 3, R.~E.~Kreiger, Melborne, FL, 1983; R.~A.~Matzner and L.~C.~Shepley, ``Classical Mechanics,'' Chapter 8.2, Prentice Hall, Englewood Cliffs, NJ, 1991.


\bibitem{FPI}
J.~J.~Halliwell,
``Derivation of the Wheeler-De Witt Equation from a Path Integral for Minisuperspace Models,''
Phys. Rev. D \textbf{38}, 2468 (1988);
J.~J.~Halliwell and J.~B.~Hartle,
``Wave functions constructed from an invariant sum over histories satisfy constraints,''
Phys. Rev. D \textbf{43}, 1170-1194 (1991);
A.~O.~Barvinsky,
``Unitarity approach to quantum cosmology,''
Phys. Rept. \textbf{230}, 237-367 (1993).

\bibitem{feng}
J.~C.~Feng and R.~A.~Matzner,
``The Weiss Variation of the Gravitational Action,''
Gen. Rel. Grav. \textbf{50}, no.8, 99 (2018)
[arXiv:1708.04489 [gr-qc]];
``From path integrals to the Wheeler-DeWitt equation: Time evolution in spacetimes with a spatial boundary,''
Phys. Rev. D \textbf{96}, no.10, 106005 (2017)
[arXiv:1708.07001 [gr-qc]];
J.~C.~Feng and S.~Chakraborty,
``Weiss variation for general boundaries,''
Gen. Rel. Grav. \textbf{54}, no.7, 67 (2022)
[arXiv:2111.06897 [gr-qc]].

\bibitem{GHY}
J.~W.~York, Jr.,
``Role of conformal three geometry in the dynamics of gravitation,''
Phys. Rev. Lett. \textbf{28}, 1082-1085 (1972);
G.~W.~Gibbons and S.~W.~Hawking,
``Action Integrals and Partition Functions in Quantum Gravity,''
Phys. Rev. D \textbf{15}, 2752-2756 (1977).


\bibitem{dileep}
J.~Chakravarty, K.~Dasgupta, D.~Jain, D.~P.~Jatkar, A.~Maji and R.~Tatar,
``Coherent states in M-theory: A brane scan using the Taub-NUT geometry,''
Phys. Rev. D \textbf{108}, no.8, L081902 (2023)
[arXiv:2308.08613 [hep-th]].


\bibitem{page}
D.~N.~Page and W.~K.~Wootters,
``Evolution without evolution: Dynamics described by stationary observables,''
Phys. Rev. D \textbf{27}, 2885 (1983)


\bibitem{witten}
E.~Witten,
``A background-independent algebra in quantum gravity,''
JHEP \textbf{03}, 077 (2024)
[arXiv:2308.03663 [hep-th]];
C.~H.~Chen and G.~Penington,
``A clock is just a way to tell the time: gravitational algebras in cosmological spacetimes,''
[arXiv:2406.02116 [hep-th]].

\bibitem{suvrat}
C.~Chowdhury, V.~Godet, O.~Papadoulaki and S.~Raju,
``Holography from the Wheeler-DeWitt equation,''
JHEP \textbf{03}, 019 (2022)
[arXiv:2107.14802 [hep-th]];
T.~Chakraborty, J.~Chakravarty, V.~Godet, P.~Paul and S.~Raju,
``The Hilbert space of de Sitter quantum gravity,''
JHEP \textbf{01}, 132 (2024)
[arXiv:2303.16315 [hep-th]]; 
``Holography of information in de Sitter space,''
JHEP \textbf{12}, 120 (2023)
[arXiv:2303.16316 [hep-th]].

\bibitem{tcc}
J.~Martin and R.~H.~Brandenberger,
``The Trans-Planckian problem of inflationary cosmology,''
Phys. Rev. D \textbf{63}, 123501 (2001)
[arXiv:hep-th/0005209 [hep-th]];
A.~Bedroya and C.~Vafa,
``Trans-Planckian Censorship and the Swampland,''
JHEP \textbf{09}, 123 (2020)
[arXiv:1909.11063 [hep-th]];
A.~Bedroya, R.~Brandenberger, M.~Loverde and C.~Vafa,
``Trans-Planckian Censorship and Inflationary Cosmology,''
Phys. Rev. D \textbf{101}, no.10, 103502 (2020)
[arXiv:1909.11106 [hep-th]];
S.~Brahma,
``Trans-Planckian censorship conjecture from the swampland distance conjecture,''
Phys. Rev. D \textbf{101}, no.4, 046013 (2020)
[arXiv:1910.12352 [hep-th]];
M.~Blamart, S.~Laliberte and R.~Brandenberger,
``TCC bounds on the static patch of de Sitter space,''
JHEP \textbf{05}, 193 (2023)
[arXiv:2301.02741 [hep-th]].

\bibitem{Dirac-algebra}
P.~A.~M.~Dirac,
``The Theory of gravitation in Hamiltonian form,''
Proc. Roy. Soc. Lond. A \textbf{246}, 333-343 (1958);
M.~Bojowald, S.~Brahma, U.~Buyukcam and F.~D'Ambrosio,
``Hypersurface-deformation algebroids and effective spacetime models,''
Phys. Rev. D \textbf{94}, no.10, 104032 (2016)
[arXiv:1610.08355 [gr-qc]];
C.~Blohmann, M.~C.~B.~Fernandes and A.~Weinstein,
``Groupoid symmetry and constraints in general relativity,''
Commun. Contemp. Math. \textbf{15}, no.01, 1250061 (2013)
[arXiv:1003.2857 [math.DG]].





\end{thebibliography}
\end{document}